\documentstyle[twocolumn]{article}
\newcommand{\cqg}[1]{{\em Class.\ Quan.\ Grav.\ }{\bf #1}}
\newcommand{\grg}[1]{{\em Gen.\ Rel.\ Grav.\ }{\bf #1}}
\newcommand{\np}[1]{{\em Nucl.\ Phys.\ }{\bf #1}}
\newcommand{\pr}[1]{{\em Phys.\ Rev.\ }{\bf #1}}
\newcommand{\prl}[1]{{\em Phys.\ Rev.\ Lett.\ }{\bf #1}}
\newcommand{\pl}[1]{{\em Phys.\ Lett.\ }{\bf #1}}
\newcommand{\jmp}[1]{{\em J. Math.\ Phys.\ }{\bf #1}}
\newcommand{\jgp}[1]{{\em J. Geom.\ Phys.\ }{\bf #1}}
\newcommand{\cmp}[1]{{\em Commun.\ Math.\ Phys.\ }{\bf #1}}
\newcommand{\mpl}[1]{{\em Mod.\ Phys.\ Lett.\ }{\bf #1}}
\newcommand{\ijmp}[1]{{\em Int.\ J. Mod.\ Phys.\ }{\bf #1}}
\newcommand{\apny}[1]{{\em Ann.\ Phys.\ (N.Y.) }{\bf #1}}
\newcommand{\ncim}[1]{{\em Nuovo Cim.\ }{\bf #1}}

\begin{document}
\onecolumn

\title{Bibliography of Publications related to \\ 
Classical and Quantum Gravity\\ 
in terms of 
Connection and Loop Variables}
\author{
       Last updated by \\
       Christopher Beetle and Alejandro Corichi \\
       Center for Gravitational Physics and Geometry \\
       Pennsylvania State University \\
       E-mail: beetle@phys.psu.edu, corichi@phys.psu.edu
}

\date{March, 1997}
\maketitle

\begin{abstract}
This bibliography attempts to give a comprehensive overview of all the
literature related to the Ashtekar connection  and
the Rovelli-Smolin loop variables.  The original version was
compiled by Peter H\"ubner in 1989, and it has been subsequently
updated by Gabriela Gonzalez, Bernd Br\"ugmann, Monica Pierri,
Troy Schilling, Alejandro Corichi and Christopher Beetle.
Information about
additional literature, new preprints, and especially corrections are
always welcome.
\end{abstract}

\newpage 
\twocolumn 

\section*{Pointers}


Here are some suggestions, intended to serve as entry points into the
literature.

First of all, for a complete and authorative presentation of canonical
gravity in the Ashtekar variables there is of course Ashtekar's latest
book [2] which appeared in 1991. The most recent general introduction
to the new variables by Ashtekar are his Les Houches lectures of 1992
[379].

The latest and up to date book in the `loop representation' is the
book by Gambini and Pullin [12], specially in lattice methods and
the `extended loop representation'. Many recent articles can be
found in the book by Ehlers and Friedrich [9].

Rather complete reviews of canonical gravity
in the Ashtekar variables
can be found in Rovelli [189], Kodama [233] and
Smolin [247].  For a critical appraisal of canonical
quantum gravity see Kucha{\v r} [277]. An overview over different
approaches to quantum gravity is given by Isham in [9].

More recent reviews on the two most prominent  viewpoints, namely
the `connection' and `loop-spin networks ' representations are 
given by Ashtekar et. al. [385]
on one side, and De Pietri and Rovelli [448] on the other. 
A {\it dialogue concerning the two chief World systems} 
is given in [487].

Finally let us mention a few more specialized references. A clear
and detailed exposition of connection dynamics is given by Romano in
[298]. For newer developments related to matter couplings (geometric
approach) see Peld\'an [369]. The definition of the loop representation is
discussed in Br\"ugmann [7]. Pullin [372] gives an introduction to
results obtained via the loop representation in (unreduced) quantum
gravity.

\newpage

\section*{Books and Dissertations}

\begin{enumerate}

\item
Abhay Ashtekar and {invited contributors}.
{\em New Perspectives in Canonical Gravity}.
Lecture Notes.  Napoli, Italy: Bibliopolis, February 1988.
[Errata published as Syracuse University preprint by Joseph D. Romano
  and Ranjeet S. Tate.]

\item
Abhay Ashtekar.
 {\em Lectures on non-perturbative canonical gravity.}
 (Notes prepared in collaboration with R. Tate).
Advanced Series in Astrophysics and Cosmology-Vol. 6. 
Singapore: World Scientific, 1991.

\item
J.C. Baez.
{\em Knots and Quantum Gravity}.
Oxford U. Press. (1994).

\item
J.C. Baez and J. Muniain.
{\em Gauge Fields, Knots, and Gravity}.
 World Scientific Press (1994).

\item
R. Borissov.
{\em Quantization of Gravity: In search of the space of
physical states}.
Ph.D. Thesis, Temple U. (1997).

\item
O. Bostr\"om. {\em Classical aspects on the road to
quantum gravity}. Ph.D. Thesis, Institute of Theoretical
Physics, G\"oteborg (1994).

\item
B. Br\"ugmann.
{\em On the constraints of quantum general relativity in the loop
representation.} 
Ph.D. Thesis, Syracuse University (May 1993)

\item 
R. Capovilla.
{\em The self-dual spin connection as the fundamental gravitational 
variable.}
Ph.D. Thesis, University of Maryland (1991).

\item
J. Ehlers and H. Friedrich, eds.
{\em Canonical Gravity: From Classical to Quantum}.
Lecture Notes in Physics 434, (Springer-Verlag, Berlin, 1995).

\item
K. Ezawa. Nonperturbative Solutions for Canonical Quantum Gravity: an
Overview. Ph.D. Thesis, Osaka U (January 1996). gr-qc/9601050.

\item
G. F\"ul\"op.
{\em Supersymmetries and Ashtekar's Variables}.
Licentiate Thesis, I.T.P. G\"oteborg (1993). 

\item
R. Gambini and J. Pullin. {\em Loops, Knots, Gauge Theory
and Quantum Gravity}. Cambridge, Cambridge University Press (1996).

\item
V. Husain. {\it Investigations on the canonical quantization of gravity.}
Ph.D. Thesis, Yale University (1989). 

\item
J. Iwasaki.
{\em On Loop-Theoretic Frameworks of Quantum Gravity}.
Ph.D. Thesis, University of Pittsburgh (April 1994).

\item
S. Koshti.
{\em Applications of the Ashtekar variables in Classical Relativity}.
Ph. D. Thesis, University of Poona (June 1991). 

\item
H.J. Matschull.
{\em Kanonishe Formulierung von Gravitations und
Supergravitations Theorien}. Ph.D. Thesis,
Hamburg University (July 1994), ISSN 0418-983.

\item
H.A, Morales-T\'ecotl.
{\em On Spacetime and Matter at Planck Lenght}.
Ph. D. Thesis SISSA/ISAS (January 1994).

\item
P. Peld\'an.
{\em From Metric to Connection: Actions for gravity, with generalizations}.
Ph.D. Thesis I.T.P. G\"oteborg (1993) ISBN 91-7032-817-X.

\item
Paul. A. Renteln. {\em Non-perturbative approaches to Quantum Gravity. }
Ph.D. Thesis, Harvard University (1988).

\item
D. Rayner.
{\em New variables in canonical quantisation and quantum gravity.}
Ph.D. Thesis, University of London (1991).

\item
J. D.  Romano.
 {\em Geometrodynamics vs. Connection Dynamics (in the context
of (2+1)- and (3+1)-gravity)}.
 Ph.D. Thesis, Syracuse University (1991), see also gr-qc/9303032

\item
V.O. Soloviev.
{\em Boundary values as Hamiltonian Variables. I. New Poisson brackets}.
Ph.D. Thesis ?????.  IHEP93-48 (submitted to J. Math. Phys.)

\item
C. Soo.
{\em Classical and quantum gravity with Ashtekar variables.}
Ph.D. Thesis, Virginia Polytechnic Institute and State University.
VPI-IHEP-92-11 (July 1992)

\item
R.S. Tate.
{\em An algebraic approach to the quantization of constrained systems:
finite dimensional examples.}
Ph.D. Thesis, Syracuse University (Aug. 1992), gr-qc/9304043

\item
T. Thiemann.
{\em On the canonical quantization of gravity in
the Ashtekar framework}. Ph.D. Thesis, Achen T.
Hochschule, 1993.

\item
J.J. Zegwaard. {\em The Loop Representation for Canonical Quantum Gravity and
its Interpretation}. Ph.D. Thesis, Utrecht University (January 1994). ISBN
90-393-0070-4.


\newpage 

\section*{Papers}

\section*{1980}

\item
Paul Sommers.
 Space spinors.
 {\em J. Math. Phys.} {\bf 21}(10):2567--2571, October 1980.

\section*{1981}

\item
Amitabha Sen.
 On the existence of neutrino ``zero-modes'' in vacuum spacetimes.
 {\em J. Math. Phys.} {\bf 22}(8):1781--1786, August 1981.

\section*{1982}

\item
Abhay Ashtekar and G.T. Horowitz.
 On the canonical approach to quantum gravity.
 {\em Phys. Rev.} {\bf D26}:3342--3353, 1982.

\item
Amitabha Sen.
 Gravity as a spin system.
 {\em Phys. Lett. } {\bf B119}:89--91, December 1982.

\section*{1984}

\item
Abhay Ashtekar.
 On the {H}amiltonian of general relativity.
 {\em Physica} {\bf A124}:51--60, 1984.

\item
A. Ashtekar and G.T. Horowitz.
Phase space of general relativity revisited: A canonical choice of
time and simplification of the Hamiltonian. \jmp{25}: 1473-1480, (1984).

\item
E.~T. Newman.
 Report of the workshop on classical and quantum alterate theories of
  gravity.
 In B.~Bertotti, F.~{de Felice}, and A.~Pascolini, editors, {\em The
  Proceedings of the 10th International Conference on General Relativity and
  Gravitation}, Amsterdam, 1984.

\section*{1986}

\item
Abhay Ashtekar.
 New variables for classical and quantum gravity.
 {\em Phys. Rev. Lett.} {\bf 57}(18):2244--2247, November 1986.

\item
Abhay Ashtekar.
 Self-duality and spinorial techniques in the canonical approach to
  quantum gravity.
 In C.~J. Isham and R.~Penrose, editors, {\em Quantum Concepts in
  Space and Time}, pages 303--317. Oxford University Press, 1986.

\item
Robert~M. Wald.
 Non-existence of dynamical perturbations of {S}chwarzschild with
  vanishing self-dual part.
 {\em Class. Quan. Grav.} {\bf 3}(1):55--63, January 1986.

\newpage

\section*{1987}

\item
Abhay Ashtekar.
 New {H}amiltonian formulation of general relativity.
 {\em Phys. Rev. } {\bf D36}(6):1587--1602, September 1987.

\item
Abhay Ashtekar.
 {E}instein constraints in the {Y}ang-{M}ills form.
 In G.~Longhi and L~Lusanna, editors, {\em Constraint's Theory and
  Relativistic Dynamics}, Singapore, 1987. World Scientific.

\item
Abhay Ashtekar, Pawel Mazur, and Charles~G. Torre.
 {BRST} structure of general relativity in terms of new variables.
 {\em Phys. Rev. } {\bf D36}(10):2955--2962, November 1987.

\item
John~L. Friedman and Ian Jack.
 Formal commutators of the gravitational constraints are not
  well-defined: A translation of {A}shtekar's ordering to the {S}chr{\"o}dinger
  representation.
 {\em Phys. Rev. } {\bf D37}(12):3495--3504, June 1987.

\item
Kazuo Ghoroku.
New variable formalism of higher derivative gravity.
\pl{B194}: 535-538, 1987

\item
Ted Jacobson and Lee Smolin.
 The left-handed spin connection as a variable for canonical gravity.
 {\em Phys. Lett. } {\bf B196}(1):39--42, September 1987.

\item
Joseph Samuel.
 A {L}agrangian basis for {A}shtekar's reformulation of canonical
  gravity.
 {\em Pram{\=a}na-J Phys.} {\bf 28}(4):L429-L432, April 1987.

\item
N.~C. Tsamis and R.~P. Woodard.
 The factor ordering problem must be regulated.
 {\em Phys. Rev.}  {\bf D36}(12):3641--3650, December 1987.

\newpage

\section*{1988}

\item
Abhay Ashtekar.
 A $3+1$ formulation of {E}instein self-duality.
 In J.~Isenberg, editor, {\em Mathematics and General Relativity},
  Providence, 1988. American Mathematical Society.

\item
Abhay Ashtekar.
 Microstructure of space-time in quantum gravity.
 In K.~C. Wali, editor, {\em Proceedings of the Eight Workshop in
  Grand Unification}, Singapore, 1988. World Scientific.

\item
Abhay Ashtekar.
 New perspectives in canonical quantum gravity.
 In B.~R. Iyer, A.~Kembhavi, J.~V. Narlikar, and C.~V. Vishveshwara,
  editors, {\em Highlights in Gravitation and Cosmology}. Cambridge University
  Press, 1988.

\item
Abhay Ashtekar, Ted Jacobson, and Lee Smolin.
 A new characterization of half-flat solutions to {E}instein's
  equation.
 {\em Commun. Math. Phys.} {\bf 115}:631--648, 1988.

\item
Ingemar Bengtsson.
 Note on {A}shtekar's variables in the spherically symmetric case.
 {\em Class. Quan. Grav.} {\bf 5}(10):L139--L142, October 1988.

\item
R. Gianvittorio, R. Gambini and A. Trias.
\pr{D38} (1988) 702

\item
J.~N. Goldberg.
 A {H}amiltonian approach to the strong gravity limit.
 {\em Gen. Rel. Grav.} {\bf 20}(9):881--891, September 1988.

\item
J.~N. Goldberg.
 Triad approach to the {H}amiltonian of general relativity.
 {\em Phys. Rev. } {\bf D37}(8):2116--2120, April 1988.

\item
Viqar Husain.
 The {$G_{\mbox{Newton}}\rightarrow\infty$} limit of quantum gravity.
 {\em Class. Quan. Grav.} {\bf 5}(4):575--582, April 1988.

\item
Ted Jacobson.
 Fermions in canonical gravity.
 {\em Class. Quan. Grav.} {\bf 5}(10):L143--L148, October 1988.

\item
Ted Jacobson.
 New variables for canonical supergravity.
 {\em Class. Quan. Grav.} {\bf 5}:923--935, 1988.

\item
Ted Jacobson.
 Superspace in the self-dual representation of quantum gravity.
 In J.~Isenberg, editor, {\em Mathematics and General Relativity},
  Providence, 1988. American Mathematical Society.

\item
Ted Jacobson and Lee Smolin.
 Covariant action for {A}shtekar's form of canonical gravity.
 {\em Class. Quan. Grav.} {\bf 5}(4):583--594, April 1988.

\item
Ted Jacobson and Lee Smolin.
 Nonperturbative quantum geometries.
 {\em Nucl. Phys.} {\bf B299}(2):295--345, April 1988.

\item
Hideo Kodama.
 Specialization of {A}shtekar's formalism to {B}ianchi cosmology.
 {\em Prog. Theor. Phys.} {\bf 80}(6):1024--1040, December 1988.

\item
Carlo Rovelli and Lee Smolin.
 Knot theory and quantum gravity.
 {\em Phys. Rev. Lett.} {\bf 61}:1155--1158, 1988.

\item
Joseph Samuel.
 Gravitational instantons from the {A}shtekar variables.
 {\em Class. Quan. Grav.} {\bf 5}:L123--L125, 1988.

\item
Lee Smolin.
 Quantum gravity in the self-dual representation.
 In J.~Isenberg, editor, {\em Mathematics and General Relativity},
  Providence, 1988. American Mathematical Society.

\item
C.~G. Torre.
 The propagation amplitude in spinorial gravity.
 {\em Class. Quan. Grav.} {\bf 5}:L63--L68, 1988.

\item
Edward Witten.
 (2+1) dimensional gravity as an exactly soluble system.
 {\em Nucl. Phys.} {\bf B311}(1):46--78, December 1988.

\newpage

\section*{1989}

\item
Abhay Ashtekar.
 Non-pertubative quantum gravity: A status report.
 In M.~Cerdonio, R.~Cianci, M.~Francaviglia, and M.~Toller, editors,
  {\em General Relativity and Gravitation}. Singapore: World Scientific, 1989.

\item
Abhay Ashtekar.
 Recent developments in {H}amiltonian gravity.
 In B.~Simon, I.~M. Davies, and A.~Truman, editors, {\em The
  Proceedings of the {IX}th International Congress on Mathematical Physics},
Swansea UK, July 1988.(Bristol, UK: Adam Hilger, 1989). 

\item
Abhay Ashtekar.
 Recent developments in quantum gravity.
 In E.~J. Fenyves, editor, {\em Proceedings of the Texas Symposium on
  Relativistic Astrophysics}. New York Academy of Science, 1989.

\item
Abhay Ashtekar.
 Recent Developments in Quantum Gravity. {\it Annals of the New York 
Academy of Sciences} {\bf 571}, 16-26. December 1989.

\item
Abhay Ashtekar, A.~P. Balachandran, and S.~G. Jo.
 The {CP}-problem in quantum gravity.
 {\em Int. Journ. Theor. Phys.} {\bf A4}:1493--1514, 1989.

\item
Abhay Ashtekar, Viqar Husain, Carlo Rovelli, Joseph Samuel, and Lee Smolin.
 $2+1$ quantum gravity as a toy model for the $3+1$ theory.
 {\em Class. Quan. Grav.} {\bf 6}:L185--L193, 1989.

\item
Abhay Ashtekar and Joseph~D. Romano.
 {C}hern-{S}imons and {P}alatini actions and ($2+1$)-gravity.
 {\em Phys. Lett. } {\bf B229}(1,2):56--60, October 1989.

\item
Abhay Ashtekar, Joseph~D. Romano, and Ranjeet~S. Tate.
 New variables for gravity: Inclusion of matter.
 {\em Phys. Rev. } {\bf D40}(8):2572--2587, October 1989.

\item
Abhay Ashtekar and Joseph~D. Romano.
 Key ($3+1$)-equations in terms of new variables (for numerical
  relativity).
 Syracuse University Report (1989).

\item
Ingemar Bengtsson.
 {Y}ang-{M}ills theory and general relativity in three and four
  dimensions.
 {\em Phys. Lett. } {\bf B220}:51--53, 1989.

\item
Ingemar Bengtsson.
 Some remarks on space-time decomposition, and degenerate metrics, in
  general relativity.
 {\em Int. J.  Mod. Phys. } {\bf A4}(20):5527--5538,
  1989.

\item
Riccardo Capovilla, John Dell, and Ted Jacobson.
 General relativity without a metric.
 {\em Phys. Rev. Lett.} {\bf 63}(21):2325--2328, November 1989.

\item
Steven Carlip.
 Exact quantum scattering in 2+1 dimensional gravity.
 {\em Nucl. Phys.} {\bf B324}(1):106--122, 1989.

\item
B. P. Dolan.
 On the generating function for Ashtekar's canonical transformation.
 {\em Phys. Lett. } {\bf B233}(1,2):89-92 , December 1989.

\item
Tevian Dray, Ravi Kulkarni, and Joseph Samuel.
 Duality and conformal structure.
 {\em J. Math. Phys.} {\bf 30}(6):1306--1309, June 1989.

\item
N.~N. Gorobey and A.~S. Lukyanenko.
 The closure of the constraint algebra of complex self-dual gravity.
 {\em Class. Quan. Grav.} {\bf 6}(11):L233--L235, November 1989.

\item
M.~Henneaux, J.~E. Nelson, and C.~Schomblond.
 Derivation of {A}shtekar variables from tetrad gravity.
 {\em Phys. Rev. } {\bf D39}(2):434--437, January 1989.

\item
A. Herdegen. Canonical gravity from a variation principle in a copy of 
a tangent bundle. {\it Class.  Quan. Grav.} {\bf 6}(8):1111-24,
(1989).

\item 
G. T. Horowitz.
Exactly soluble diffeomorphism invariant theories. {\it Commun. 
Math. Phys.} {\bf 125}(3): 417-37, 1989.

\item
Viqar Husain.
 Intersecting loop solutions of the {H}amiltonian constraint of
  quantum general relativity.
 {\em Nucl. Phys.} {\bf B313}:711--724, 1989.

\item
Viqar Husain and Lee Smolin.
 Exactly solvable quantum cosmologies from two {K}illing field
  reductions of general relativity.
 {\em Nucl. Phys.} {\bf B327}:205--238, 1989.

\item
V.~Khatsymovsky.
 Tetrad and self-dual formulation of {R}egge calculus.
 {\em Class. Quan. Grav.} {\bf 6}(12):L249--L255, December 1989.

\item
Sucheta Koshti and Naresh Dadhich.
 Degenerate spherical symmetric cosmological solutions using
  {A}shtekar's variables.
 {\em Class. Quan. Grav.} {\bf 6}:L223--L226, 1989.

\item
Stephen~P. Martin.
 Observables in 2+1 dimensional gravity.
 {\em Nucl. Phys.} {\bf 327}(1):78--204, November 1989.

\item
L.~J. Mason and E.~T. Newman.
 A connection between {E}instein and {Y}ang-{M}ills equations.
 {\em Commun. Math. Phys.} {\bf 121}(4):659--668, 1989.

\item
J.~E. Nelson and T.~Regge.
 Group manifold derivation of canonical theories.
 {\em Int. J. Mod. Phys.} {\bf A4},2021 (1989).

\item
Paul Renteln and Lee Smolin.
 A lattice approach to spinorial quantum gravity.
 {\em Class. Quan. Grav.} {\bf 6}:275--294, 1989.

\item
Amitabha Sen and Sharon Butler.
 The quantum loop.
 {\em The Sciences}: 32--36, November/December 1989.

\item
L. Smolin.
 Invariants of links and critical points of the {C}hern-{S}imon path
  integrals.
 {\em Mod. Phys. Lett.} {\bf A4}:1091--1112, 1989.

\item
L. Smolin.
Loop representation for quantum gravity in 2+1 dimensions.
In the {\em Proceedings of the John's Hopkins Conference on Knots,
Topology and Quantum Field Theory}, ed. L. Lusanna (World Scientific,
Singapore 1989)

\item
Sanjay~M. Wagh and Ravi~V. Saraykar.
 Conformally flat initial data for general relativity in {A}shtekar's
  variables.
 {\em Phys. Rev. } {\bf D39}(2):670--672, January 1989.

\item
Edward Witten.
 Gauge theories and integrable lattice models.
 {\em Nucl. Phys.} {\bf B322}(3):629--697, August 1989.

\item
Edward Witten.
 Topology-changing amplitudes in (2+1) dimensional gravity.
 {\em Nucl. Phys.} {\bf B323}(1):113--122, August 1989.

\newpage

\section*{1990}

\item
C. Aragone and A. Khouder .
 Vielbein gravity in the light-front gauge.
 {\em Class. Quan. Grav.} {\bf 7}:1291--1298, 1990.

\item
Abhay Ashtekar.
Old problems in the light of new variables.
In {\em Proceedings of the Osgood Hill Conference on Conceptual
Problems in Quantum Gravity}, eds. A. Ashtekar and J. Stachel
(Birkh\"auser, Boston 1991)

\item
Abhay Ashtekar.
 Self duality, quantum gravity, {W}ilson loops and all that.
 In N.~Ashby, D.~F. Bartlett, and W.~Wyss, editors, {\em Proceedings
  of the 12th International Conference on General Relativity and Gravitation}.
  Cambridge University Press, 1990.

\item
Abhay Ashtekar and Jorge Pullin.
 {B}ianchi cosmologies: A new description.
 {\em Proc. Phys. Soc. Israel} {\bf 9}:65-76 (1990).

\item
Abhay Ashtekar.
 Lessons from 2+1 dimensional quantum gravity.
In {\em "Strings 90"} edited
by R. Arnowitt et al (Singapore: World Scientific, 1990).

\item
Ingemar Bengtsson.
 A new phase for general relativity?
 {\em Class. Quan. Grav.} {\bf 7}(1):27--39, January 1990.

\item
Ingemar Bengtsson.
 P, T, and the cosmological constant.
 {\em Int. J.  Mod. Phys. } {\bf A5}(17):3449-3459 (1990).

\item
Ingemar Bengtsson.
 Self-Dual Yang-Mills fields and Ashtekar variables.
 {\em Class. Quan. Grav.} {\bf 7}:L223-L228 (1990)

\item
Ingemar Bengtsson and P. Peld{\' a}n.
Ashtekar variables, the theta-term, and the cosmological constant.
{\em Phys. Lett.} {\bf B244}(2): 261-64, 1990.

\item
M.~P. Blencowe.
 The {H}amiltonian constraint in quantum gravity.
 {\em Nuc. Phys.} {\bf B341}(1):213, 1990.

\item
L.~Bombelli and R.~J. Torrence.
 Perfect fluids and {A}shtekar variables, with applications to
  {K}antowski-{S}achs models.
 {\em Class.Quan. Grav.} {\bf 7}:1747 (1990).

\item
Riccardo Capovilla, John Dell, and Ted Jacobson.
 Gravitational instantons as {SU(2)} gauge fields.
 {\em Class.Quan. Grav.} {\bf 7}(1):L1--L3, January 1990.

\item
Steven Carlip.
 Observables, gauge invariance and time in 2+1 dimensional gravity.
 {\em Phys. Rev.} {\bf D42}, 2647-2654 (October 1990).

\item
S. Carlip and S. P. de Alwis.
 Wormholes in (2+1)-gravity.
 {\em Nuc. Phys.} {\bf B337}:681-694, June 1990.

\item
G. Chapline.
Superstrings and Quantum Gravity.
{\em Mod. Phys. Lett.}{\bf A5}:2165-72 (1990).

\item
R. Floreanini and R. Percacci.
 Canonical algebra of GL(4)-invariant gravity.
 {\em Class.Quan. Grav.} {\bf 7}:975--984, 1990.

\item
R. Floreanini and R. Percacci.
 Palatini formalism and new canonical variables for GL(4)-invariant
gravity.
{\em Class. Quan. Grav.} {\bf 7}: 1805-18, 1990.

\item
R. Floreanini and R. Percacci.
 Topological pregeometry.
 {\em Mod. Phys. Lett.} {\bf A5}: 2247-51, 1990.

\item
Takeshi Fukuyama and Kiyoshi Kaminura.
 Complex action and quantum gravity.
{\em Phys. Rev.} {\bf D41}:1105-11, February 1990.

\item
G.~Gonzalez and J.~Pullin.
 {BRST} quantization of 2+1 gravity.
 {\em Phys. Rev. } {\bf D42}(10): 3395-3400 (1990).
[Erratum: {\em Phys. Rev.} {\bf 43}: 2749, April 1991].

\item
N.~N. Gorobey and A.~S. Lukyanenko.
 The {A}shtekar complex canonical transformation for supergravity.
 {\em Class. Quan. Grav.} {\bf 7}(1):67--71, January 1990.

\item
C. Holm.
 Connections in Bergmann manifolds.
 {\em Int. Journ. Theor. Phys.} {\bf A29}(1):23-36, January 1990.

\item
V. Husain and K. Kucha{\v r}.
 General covariance, the New variables, and dynamics without dynamics.
{\em Phys. Rev.} {\bf D42}(12)4070-4077 (December 1990).

\item
Viqar Husain and Jorge Pullin.
 Quantum theory of space-times with one Killing field.
 {\em Modern Phys. Lett. } {\bf A5}(10):733-741, April 1990.

\item
K. Kamimura and T. Fukuyama. Ashtekar's formalism in 1st order tetrad form.
{\em Phys. Rev.}  {\bf D41}(6): 1885-88, 1990. 

\item
H. Kodama.
 Holomorphic wavefunction of the universe.
 {\em Phys. Rev.} {\bf D42}: 2548-2565 (October 1990).

\item
Sucheta Koshti and Naresh Dadhich.
 On the self-duality of the {W}eyl tensor using {A}shtekar's
  variables.
 {\em Class. Quan. Grav.} {\bf 7}(1):L5--L7, January 1990.

\item
Noah Linden.
 New designs on space-time foams.
 {\em Physics World} {\bf 3}(3):30-31, March 1990.

\item
N.Manojlovic.
 Alternative loop variables for canonical gravity.
 {\em Class. Quan. Grav.} {\bf 7}:1633-1645. (1990).

\item
E.~W. Mielke.
 Generating functional for new variables in general relativity and
  {P}oincare gauge theory.
 {\em Phys. Lett.} {\bf A149}: 345-350 (1990).

\item
E.~W. Mielke.
 Positive gravitational energy proof from complex variables?
 {\em Phys. Rev.} {\bf D42}(10): 3338-3394 (1990).

\item
Peter Peld\'{a}n.
 Gravity coupled to matter without the metric.
{\em Phys. Lett.} {\bf B248}(1,2): 62-66 (1990).

\item
D.~Rayner.
 A formalism for quantising general relativity using non-local
  variables.
 {\em Class. Quan. Grav.} {\bf 7}(1):111--134, January 1990.

\item
D.~Rayner.
Hermitian operators on quantum general relativity loop space.
 {\em Class. Quan. Grav.} {\bf 7}(4):651--661, April 1990.

\item
Paul Renteln.
 Some results of {SU}(2) spinorial lattice gravity.
 {\em Class. Quan. Grav.} {\bf 7}(3):493--502, March 1990.

\item
D.C. Robinson and C. Soteriou.
 Ashtekar's new variables and the vacuum constraint equations.
 {\em Class. Quan. Grav.} {\bf 7}(11): L247-L250 (1990).

\item
Carlo Rovelli and Lee Smolin.
 Loop representation of quantum general relativity.
 {\em Nuc. Phys.} {\bf B331}(1): 80-152, February 1990.

\item
M.~Seriu and H.~Kodama.
 New canonical formulation of the {E}instein theory.
 {\em Prog. Theor. Phys.} {\bf 83}(1):7-12, January 1990.

\item
Lee Smolin.
 Loop representation for quantum gravity in $2+1$ dimensions.
 In {\em Proceedings of the 12th John Hopkins Workshop: Topology and
  Quantum Field Theory} (Florence, Italy), 1990.

\item
C. G. Torre.
 Perturbations of gravitational instantons.
 {\em Phys. Rev.} {\bf D41}(12) : 3620-3621, June 1990.

\item
C.~G. Torre.
 A topological field theory of gravitational instantons.
 {\em Phys. Lett } {\bf B252}(2):242-246 (1990).

\item
C. G. Torre.
 On the linearization stability of the conformally
(anti)self dual  {E}instein equations,
 {\em J. Math. Phys.} {\bf 31}(12): 2983-2986 (1990).

\item
H. Waelbroeck.
 2+1 lattice gravity.
 {\em Class. Quan. Grav.} {\bf 7}(1): 751--769, January 1990.

\item
M. Waldrop.
Viewing the Universe as a Coat of Chain Mail. 
{\em Science} {\bf 250}: 1510-1511 (1990).

\item
R. P. Wallner
 New variables in gravity theories.
 {\em Phys. Rev. } {\bf D42}(2):441-448 ,July  1990.

\item
R.S. Ward.
 The SU($\infty$) chiral model and self-dual vacuum spaces.
 {\em Class. Quan. Grav.} {\bf 7}: L217-L222 (1990).

\newpage

\section*{1991}

\item
V. Aldaya and J. Navarro-Salas.
New solutions of the hamiltonian and diffeomorphism constraints of quantum 
gravity from a highest weight loop representation.
{\em Phys. Lett.} {\bf B259}: 249-55, April 1991.

\item
Abhay Ashtekar.
Old problems in the light of new variables.
In {\em Proceedings of the Osgood Hill Conference on Conceptual
Problems in Quantum Gravity}, eds. A. Ashtekar and J. Stachel
(Birkh\"auser, Boston 1991)

\item
Abhay Ashtekar.
The winding road to quantum gravity.
In {\em Proceedings of the Osgood Hill Conference on Conceptual
Problems in Quantum Gravity}, eds. A. Ashtekar and J. Stachel
(Birkh\"auser, Boston 1991)

\item
Abhay Ashtekar.
Canonical Quantum Gravity.
In {\em The Proceedings of the 1990 Banff Workshop on Gravitational
Physics}, edited by R. Mann (Singapore: World Scientific, 1991), and
in the {\em Proceedings of SILARG VIII Conference}, edited by M.
Rosenbaum and M. Ryan (Singapore: World Scientific 1991).

\item
A. Ashtekar, C. Rovelli and L. Smolin.
Gravitons and loops.
{\em Phys. Rev.}{\bf D44}(6):1740-55, 15 September 1991.

\item
A. Ashtekar and J. Samuel.
Bianchi cosmologies: the role of spatial topology.
\cqg{8} (1991) 2191--215

\item
I. Bengtsson.
 The cosmological constants.
{\em Phys. Lett.} {\bf B254}:55-60, 1991.

\item
I. Bengtsson.
Self-duality and the metric in a family of neighbours of Einstein's equations.
{\em J. Math. Phys.}{\bf 32} (Nov. 1991) 3158--61

\item
I. Bengtsson.
Degenerate metrics and an empty black hole. 
\cqg{8}, 1847 (1991),
Goteborg-90-45 (December 1990).

\item
Peter G. Bergmann and Garrit Smith.
Complex phase spaces and complex gauge groups in general relativity.
 {\em Phys. Rev.} {\bf D43}:1157-61,  February 1991.

\item
L. Bombelli.
Unimodular relativity, general covariance, time, and the Ashtekar
variables.
In {\em Gravitation. A Banff Summer Institute}, eds.~R. Mann and P.
Wesson (World Scientific 1991) 221--32

\item
L. Bombelli, W.E. Couch and R.J.Torrence.
Time as spacetime four-volume and the Ashtekar variables.
\pr{D44} (15. Oct. 1991) 2589--92

\item
B. Br\"{u}gmann.
 The method of loops applied to lattice gauge theory.
{\em Phys. Rev. } {\bf D43}: 566-79, January 1991.

\item
B. Br\"{u}gmann and J. Pullin.
 Intersecting N loop solutions of the Hamiltonian constraint
of Quantum Gravity.
 {\em Nuc. Phys.} {\bf B363}: 221-44, September 1991.

\item
R. Capovilla, J. Dell, T. Jacobson and L. Mason.
 Self dual forms and gravity.
{\em Class. Quan. Grav.} {\bf 8}: 41-57, January 1991.

\item
R. Capovilla, J. Dell and T. Jacobson.
 A pure spin-connection formulation of gravity.
{\em Class. Quan. Grav.} {\bf 8}: 59-74, January 1991.

\item
Steven Carlip.
 Measuring the metric in 2+1 dimensional quantum gravity.
{\em Class. Quan. Grav.} {\bf 8}:5-17, January 1991.

\item
S. Carlip and J. Gegenberg.
Gravitating topological matter in 2+1 dimensions. 
{\em Phys. Rev.}{\bf D44}(2):424-28, 15 July 1991.

\item
L. Crane.
2-d physics and 3-d topology. 
{\em Commun. Math. Phys.} {\bf 135}: 615-640, January 1991.

\item
N. Dadhich, S. Koshti and A. Kshirsagar.
 On constraints of pure connection formulation of General Relativity
for non-zero cosmological constant.
{\em Class. Quan. Grav.} {\bf 8}: L61-L64, March 1991.

\item
B.~P. Dolan.
 The extension of chiral gravity to {SL}(2,{C}).
In {\em Proceedings of the
1990 Banff Summer School on gravitation}, ed. by R. Mann (World Scientific,
Singapore 1991)

\item
R. Floreanini and R. Percacci.
 GL(3) invariant gravity without metric.
 {\em Class. Quan. Grav.}{\bf 8}(2):273-78, February 1991.

\item
G. Fodor and Z. Perjes.
Ashtekar variables without hypersurfaces.
{\em Proc. of Fifth Sem. Quantum Gravity, Moscow} (Singapore: World
Scientific 1991) 183--7

\item
H. Fort and R. Gambini.
Lattice QED with light fermions in the P representation.
IFFI preprint, 90-08. \pr{D44}:1257-1262, 1991.

\item
T. Fukuyama and K. Kamimura.
Schwarzschild solution in Ashtekar formalism.
\mpl{A6} (1991) 1437--42

\item
R. Gambini.
Loop space representation of quantum general relativity and the group of loops.
{\em Phys. Lett.} {\bf B255}:180-88, February 1991.

\item
J.N. Goldberg.
Self-dual Maxwell field on a null cone.
\grg{23} (December 1991) 1403--1413 

\item
J.N. Goldberg, E.T. Newman, and C. Rovelli.
On Hamiltonian systems with first class constraints.
\jmp{32}(10) (1991) 2739--43

\item
J. Goldberg, D.C. Robinson and C. Soteriou.
Null surface canonical formalism. In {\em  Gravitation and Modern Cosmology},
ed. Zichichi (Plenum Press, New York, 1991)

\item
J. Goldberg, D.C.Robinson and C. Soteriou.  
A canonical formalism with a self-dual Maxwell field on a null surface.  
In {\em 9th Italian Conference on General Relativity and Gravitational
Physics (P.G.  Bergmann Festschrift)}, ed. R. Cianci et al (World
Scientific, Singapore 1991)

\item
G.T. Horowitz.
Topology change in classical and quantum gravity.
{\em Class. Quan. Grav.} {\bf 8}:587-601, April 1991.

\item
V. Husain.
Topological quantum mechanics.
{\em Phys. Rev.} {\bf D43}: 1803-07, March 1991.

\item
H. Ikemori.
Introduction to two form gravity and Ashtekar formalism.
YITP-K-922 preprint (March 1991).
{\em Tokyo Quantum Gravity}:7-88, 1991.

\item
C. J. Isham.
Loop Algebras and Canonical Quantum Gravity.
To appear in Contemporary Mathematics,edited by M. Gotay, V. Moncrief 
and J. Marsden (American Mathematical Society, Providence, 1991).

\item
K. Kamimura, S. Makita and T. Fukuyama .
Spherically symmetric vacuum solution in Ashtekar's formulation of gravity.
\mpl{A6} (30. Oct. 1991) 3047--53

\item
C. Kozameh and E.T. Newman.
The O(3,1) Yang-mills equations and the Einstein equations.
{\em Gen. Rel. Grav.} {\bf 23}:87-98, January 1991. 

\item
H. C. Lee and Z. Y. Zhu.
Quantum holonomy and link invariants.
{\em Phys. Rev.}{\bf D44}(4):R942-45, 15 August 1991. 

\item
R. Loll.
 A new quantum representation for canonical gravity and SU(2)
Yang-Mills theory.
\np{B350} (1991) 831--60

\item
E. Mielke, F. Hehl.
Comment on ``General relativity without the metric''.
\prl{67} (Sept.\ 1991) 1370

\item
V. Moncrief and M. P. Ryan.
Amplitude-real-phase exact solutions for quantum mixmaster universes.
{\em Phys. Rev.}{\bf D44}, (1991), 2375.

\item
C.~Nayak.
 The loop space representation of 2+1 quantum gravity: physical
  observables,variational principles,  and the issue of time.
{\em Gen. Rel. Grav.} {\bf 23}: 661-70, June 1991.

\item
C.~Nayak.
Einstein-Maxwell theory in 2+1 dimensions.
{\em Gen. Rel. Grav.}{\bf 23}:981-90, September 1991.

\item
H. Nicolai.
The canonical structure of maximally extended supergravity in three dimensions.
\np{B353} (April 1991) 493

\item
P. Peld{\' a}n.
Legendre transforms in Ashtekar's theory of gravity.
\cqg{8} (Oct. 1991) 1765--83

\item
P. Peld{\' a}n.
Non-uniqueness of the ADM Hamiltonian for gravity.
\cqg{8} (Nov. 1991) L223--7

\item
C. Rovelli.
Ashtekar's formulation of general relativity and loop-space non-perturbative
quantum gravity : a report. 
{\em Class. Quan. Grav.}{\bf 8}(9): 1613-1675, September 1991.

\item
Carlo Rovelli.
 Holonomies and loop representation in quantum gravity.
In {\em The Newman Festschrift}, ed. by A. Janis and J. Porter.
(Birkh{\" a}user, Boston 1991)

\item
Joseph Samuel.
 Self-duality in Classical Gravity.
In {\em The Newman Festschrift}, ed. by A. Janis and J. Porter.
(Birkh{\" a}user, Boston 1991)

\item
Lee Smolin.
 Nonperturbative quantum gravity via the loop representation.
In {\em Conceptual
Problems of Quantum Gravity}, eds. A. Ashtekar and J. Stachel
(Birkh\"auser, Boston, 1991)

\item
G. t'Hooft.
A chiral alternative to the vierbein field in general relativity.
\np{B357}: 211-221, 1991.

\item
S. Uehara.
A note on gravitational and SU(2) instantons with Ashtekar variables.
\cqg{8} (Nov. 1991) L229--34

\item
M. Varadarajan.
Non-singular degenerate negative energy solution to the Ashtekar equations.
\cqg{8} (Nov. 1991) L235--40

\item
K. Yamagishi and G.F. Chapline.
Induced 4-d self-dual quantum gravity: $\hat{W}_{\infty}$ algebraic approach.
{\em Class. Quan. Grav.}{\bf 8}(3):427-46, March 1991.

\item
J. Zegwaard.
Representations of quantum general relativity using Ashtekar's variables.
{\em Class. Quan. Grav.}{\bf 8} (July 1991) 1327--37

\newpage

\section*{1992}

\item
A. Ashtekar.
Loops, gauge fields and gravity.
In {\em Proceedings of the VIth Marcel Grossmann meeting on general
relativity}, eds.\ H. Sato and T. Nakamura (World Scientific, 1992),
and in {\em Proceedings of the VIIIth Canadian conference on
general relativity and gravitation}, edited by G. Kunstater et al
(World Scientific, Singapore 1992)

\item
A. Ashtekar and C. Isham.
Representations of the holonomy algebras of gravity and non-abelian
gauge theories.
\cqg{9} (June 1992) 1433--85

\item
A. Ashtekar and C. Isham.
Inequivalent observable algebras: a new ambiguity in field
quantisation.
\pl{B274} (1992) 393--398

\item
A. Ashtekar and J.D. Romano.
Spatial infinity as a boundary of space-time.
\cqg{9} (April 1992) 1069--100

\item
A. Ashtekar and C. Rovelli. 
Connections, loops and quantum general relativity.
\cqg{9} suppl. (1992) S3--12

\item
A. Ashtekar and C. Rovelli.
A loop representation for the quantum Maxwell field. 
\cqg{9} (May 1992) 1121--50

\item
A. Ashtekar, C. Rovelli and L. Smolin.
Self duality and quantization.
{\em J. Geom. Phys.} {\bf 8} (1992) 7--27

\item
A. Ashtekar, C. Rovelli and L. Smolin.
Weaving a classical geometry with quantum threads.
\prl{69} (1992) 237--40

\item
J.C. Baez.
Link invariants of finite type and perturbation theory.
Lett. Math. Phys {\bf 26} (1992) 43--51.

\item
I. Bengtsson and O. Bostr\"om.
Infinitely many cosmological constants.
\cqg{9} (April 1992) L47--51

\item
I. Bengtsson and P. Peldan.
Another `cosmological' constant.
\ijmp{A7} (10 March 1992) 1287--308

\item
O. Bostr\"om.
Degeneracy in loop variables; some further results.
\cqg{9} (Aug. 1992) L83--86

\item
B. Br\"{u}gmann, R. Gambini and J. Pullin. 
Knot invariants as nondegenerate quantum geometries.
\prl{68} (27 Jan. 1992) 431--4

\item
B. Br\"{u}gmann, R. Gambini and J. Pullin. 
Knot invariants as nondegenerate states of four-dimensional quantum gravity.
In {\em Proceedings of the Twentieth International conference on 
Differential Geometric Methods in Theoretical Physics, Baruch College, City 
University of New York,1-7 June 1991}, S. Catto, A. Rocha eds. (World
Scientific, Singapore 1992)

\item
B. Br\"{u}gmann, R. Gambini and J. Pullin.
Jones polynomials for intersecting knots as physical states of quantum
gravity.  
\np{B385} (Oct.\ 1992) 587--603

\item
R. Capovilla.
Nonminimally coupled scalar field and Ashtekar variables.
\pr{D46} (Aug. 1992) 1450--

\item
R. Capovilla.
 Generally covariant gauge theories.
 UMDGR 90-253 Preprint, May 1990, \np{B373}: 233-246, (1992).

\item
R. Capovilla and T. Jacobson.
Remarks on pure spin connection formulation of gravity.
Maryland preprint UMDGR-91-134, {\em Mod. Phys. Lett} {\bf A7}: 1871-1877,
(1992). 

\item
L. N. Chang and C. P. Soo.
Ashtekar's Variables and the Topological Phase of Quantum Gravity.
In {\em Proceedings of the Twentieth International conference on 
Differential Geometric Methods in Theoretical Physics, Baruch College, City 
University of New York,1-7 June 1991}, S. Catto, A. Rocha eds. (World
Scientific, Singapore 1992)

\item
L. N. Chang and C. P. Soo.
BRST cohomology and invariants of four-dimensional gravity in
Ashtekar's variables. 
\pr{D46} (Nov. 1992) 4257--62

\item
S. Carlip.
(2+1)-dimensional Chern-Simons gravity as a Dirac square root.
\pr{D45} (1992) 3584--90

\item
Y.M. Cho, K.S. Soh, J.H. Yoon and Q.H. Park.
Gravitation as gauge theory of diffeomorphism group.
\pl{B286}: 251-255, 1992.

\item
G. F\"ul\"op.
Transformations and BRST-charges in 2+1 dimensional gravitation.
gr-qc/9209003, \mpl{A7} (1992) 3495--3502

\item
T. Fukuyama.
Exact Solutions in Ashtekar Formalism.
In {\em Proceedings of the VIth Marcel Grossmann meeting on
general relativity}, eds.\ H. Sato and T. Nakamura (World Scientific, 1992)

\item 
T. Fukuyama, K. Kamimura and S. Makita.
Metric from non-metric action of gravity.
\ijmp{D1} (1992) 363--70

\item
A. Giannopoulos and V. Daftardar.
The direct evaluation of the Ashtekar variables for any given metric
using the algebraic computing system STENSOR.
\cqg{9} (July 1992) 1813--22 

\item
J. Goldberg.
Quantized self-dual Maxwell field on a null surface.
\jgp{8} (1992) 163--172

\item
J. Goldberg.
Ashtekar variables on null surfaces. 
In {\em Proceedings of the VIth Marcel Grossmann meeting on
general relativity}, eds.\ H. Sato and T. Nakamura (World Scientific, 1992)

\item
J.N. Goldberg, J. Lewandowski, and C. Stornaiolo.
Degeneracy in loop variables.
\cmp{148} (1992) 377--402 

\item
J.N. Goldberg, D.C. Robinson and C. Soteriou.
Null hypersurfaces and new variables. 
\cqg{9} (May 1992) 1309--28

\item 
J. Horgan.
Gravity quantized?
{\em Scientific American} (Sept.\ 1992) 18--20

\item
V. Husain.
2+1 gravity without dynamics.
\cqg{9} (March 1992) L33--36 

\item
T. Jacobson and J.D. Romano.
Degenerate Extensions of general relativity.
\cqg{9} (Sept. 1992) L119--24

\item
A. Kheyfets and W.A. Miller.
E. Cartan moment of rotation in Ashtekar's 
self-dual representation of gravitation.
\jmp{33} (June 1992) 2242--

\item
C. Kim, T. Shimizu and K. Yushida.
2+1 gravity with spinor field.
\cqg{9} (1992) 1211-16

\item 
H. Kodama.
Quantum gravity by the complex canonical formulation.
gr-qc/9211022, \ijmp{D1} (1992) 439 

\item
S. Koshti.
Massless Einstein Klein-Gordon equations in the spin connection formulation.
\cqg{9} (1992) 1937--42

\item
J. Lewandowski.
Reduced holonomy group and Einstein's equations with a cosmological
constant.
\cqg{9} (Oct. 1992) L147--51 

\item
R. Loll.
Independent SU(2)-loop variables and the reduced configuration space
of SU(2)-lattice gauge theory.
\np{B368} (1992) 121--42

\item
R. Loll.
Loop approaches to gauge field theory.
Syracuse SU-GP-92/6-2, in {\em Memorial
Volume for M.K. Polivanov, Teor. Mat. Fiz.} {\bf 91} (1992)

\item
A. Magnon. 
Ashtekar variables and unification of gravitational and
electromagnetic interactions.
\cqg{9} suppl. (1992) S169--81

\item
J. Maluf.
Self-dual connections, torsion and Ashtekar's variables.
\jmp{33} (Aug.\ 1992) 2849--54

\item
J.W. Maluf.
Symmetry properties of Ashtekar's formulation of canonical gravity.
Nuovo Cimento {\bf 107} (July 1992) 755--

\item
N. Manojlovi{\' c} and A. Mikovi{\' c}.
Gauge fixing and independent canonical variables in the Ashtekar formalism
of general relativity.
\np{B382} (June 1992) 148--70

\item
N. Manojlovi{\' c} and A. Mikovi{\' c}.
Ashtekar Formulation of (2+1)-gravity on a torus.
\np{B385} (July 1992) 571--586

\item 
E.W. Mielke.
Ashtekar's complex variables in general relativity and its teleparallelism 
equivalent. 
\apny{219} (1992) 78--108

\item
E.T. Newman and C. Rovelli.
Generalized lines of force as the gauge-invariant degrees of freedom
for general relativity and Yang-Mills theory.
\prl{69} (1992) 1300--3

\item
P. Peld\'an.
Connection formulation of (2+1)-dimensional Einstein gravity and
topologically massive gravity. 
\cqg{9} (Sept. 1992) 2079--92

\item
L. Smolin.
The ${\rm G_{Newton}\rightarrow 0}$ limit of Euclidean quantum gravity.
\cqg{9} (April 1992) 883--93

\item
L. Smolin.
Recent developments in nonperturbative quantum gravity.
In {\em Proceedings of the XXII Gift International Seminar
on Theoretical Physics, Quantum Gravity and Cosmology, June 1991,
Catalonia, Spain} (World Scientific, Singapore 1992)

\item
V. Soloviev.
Surface terms in Poincare algebra in Ashtekar's formalism.
In {\em Proceedings of the VIth Marcel Grossmann meeting on
general relativity}, eds.\ H. Sato and T. Nakamura (World Scientific, 1992)

\item
Vladimir Soloviev.
How canonical are Ashtekar's variables?
\pl{B292}:30, 1992.

\item
R.S. Tate.
Polynomial constraints for general relativity using real
geometrodynamical variables. 
\cqg{9} (Jan. 1992) 101--19

\item
C.G. Torre.
Covariant phase space formulation of parametrized field theory.
\jmp{33} (Nov. 1992) 3802--12

\item
R.P. Wallner.
Ashtekar's variables reexamined.
\pr{D46} (Nov. 1992) 4263--4285

\item
J. Zegwaard.
Gravitons in loop quantum gravity.
\np{B378} (July 1992) 288--308

\newpage

\section*{1993}

\item
A. Ashtekar.
Recent developments in classical and quantum theories of connections
including general relativity.
In {\em Advances in Gravitation and Cosmology}, eds.\ B. Iyer, A.
Prasanna, R. Varma and C. Vishveshwara (Wiley Eastern, New Delhi 1993)

\item
A. Ashtekar, lecture notes by R.S. Tate.
Physics in loop space.
In {\em Quantum gravity, gravitational radiation and large scale
structure in the universe}, eds. B.R.\ Iyer, S.V. Dhurandhar and K.
Babu Joseph (1993)

\item
A. Ashtekar and J. Lewandowski.
Completeness of Wilson loop functionals on the moduli space of
$SL(2,C)$ and $SU(1,1)$-connections.
gr-qc/9304044, \cqg{10} (June 1993) L69--74

\item
A. Ashtekar, R.S. Tate and C. Uggla.
Minisuperspaces: observables, quantization and singularities.
Int. J. Mod. Phys. {\bf D2}, 15--50 (1993).

\item
A. Ashtekar, R.S. Tate and C. Uggla.
Minisuperspaces: symmetries and quantization.
In {\em Misner Festschrift}, edited by B.L. Hu, M. Ryan and C.V.
Vishveshwara (Cambridge University Press, 1993)

\item
J.C. Baez.
Quantum gravity and the algebra of tangles.
\cqg{10} (April 1993) 673--94

\item
I. Bengtsson.
Some observations on degenerate metrics.
\grg{25} (Jan. 1993) 101--12

\item
I. Bengtsson.
Strange Reality:  Ashtekar's variables with Variations.
Theor. Math. Phys. {\bf 95} (May 1993) 511

\item
I. Bengtsson.
Neighbors of Einstein's equations --- some new results.
G\"oteborg preprint ITP92-35
\cqg{10}: 1791-1802, 1993.

\item
J. Birman.
New points of view in knot theory.
{\em Bull. AMS}{\bf 28} (April 1993) 253--287

\item
B. Br\"ugmann, R. Gambini and J. Pullin.
How the Jones polynomial gives rise to physical states of quantum
general relativity.
\grg{25} (Jan.\ 1993) 1--6

\item
B. Br\"{u}gmann and J. Pullin. 
On the constraints of quantum gravity in the loop representation.
\np{B390} (Feb.\ 1993) 399--438

\item
R. Capovilla and Jerzy Pleba\'nski.
Some exact solutions of the Einstein field equations in terms of the
self-dual spin connection.
\jmp{34} (Jan.\ 1993) 130--138

\item
S. Carlip.
Six ways to quantize (2+1)-dimensional gravity.
gr-qc/9305020, {\em Canadian Gen. Rel.} (1993), 215

\item
C. Di Bartolo, R. Gambini and J. Griego.
The Extended Loop Group:  An Infinite Dimensional Manifold
Associated With the Loop Space.IFFI/93.01, gr-qc/9303010
\cmp{158} (Nov. 1993) 217--40

\item
R. Gambini and J. Pullin.
Quantum Einstein-Maxwell fields: a unified viewpoint from the loop
representation. 
hep-th/9210110, \pr{D47} (June 1993) R5214--8

\item
D.E. Grant.
On self-dual gravity.
gr-qc/9301014, \pr{d48} (Sept. 1993) 2606--12

\item
V. Husain.
Ashtekar variables, self-dual metrics and $W_\infty$.
\cqg{10} (March 1993) 543--50

\item
V. Husain.
General covariance, loops, and matter.
gr-qc/9304010, \pr{D47} (June 1993) 5394--9

\item
V. Husain.
Faraday Lines and Observables for the Einstein-Maxwell Theory.
\cqg{10} (1993) L233--L237

\item
G. Immirzi.
The reality conditions for the new canonical variables of general relativity.
\cqg{10} (Nov. 1993) 2347--52

\item
T. Jacobson and J.D. Romano.
The Spin Holonomy Group in General Relativity.
\cmp{155} (July 1993) 261--76

\item
C. Kiefer.
Topology, decoherence, and semiclassical gravity.
gr-qc/9306016, \pr{D47} (June 1993) 5413--21

\item
K. Kuchar. Canonical quantum gravity.
gr-qc/9304012. In {\em General Relativity and Cosmology 1992},
R.J Gleiser, C Kozameh, O.M. Moreshi eds. (IOP Publishing, 1993).

\item
H. Kunitomo and T. Sano.
The Ashtekar formulation for canonical $N=2$ supergravity.
Prog. Theor. Phys. suppl. (1993) 31

\item
K. Kamimura and T. Fukuyama.
Massive analogue of Ashtekar-CDJ action.
Vistas in astronomy {\bf 37} (1993) 625--

\item
S. Lau.
Canonical variables and quasilocal energy in general relativity.
gr-qc/9307026, \cqg{10} (Nov. 1993) 2379--99

\item
H.Y. Lee, A. Nakamichi and T. Ueno.
Topological two-form gravity in four dimensions.
\pr{D47} (Feb.\ 1993) 1563--68 

\item
J. Lewandowski.
Group of loops, holonomy maps, path bundle and path connection.
\cqg{10} (1993) 879--904

\item
R. Loll.
Lattice gauge theory in terms of independent Wilson loops.
In {\em Lattice 92}, eds J. Smit and P. van Baal, \np{B}
(Proc.\ Suppl.) {\bf 30} (March 1993)

\item
R. Loll.
Loop variable inequalities in gravity and gauge theory.
\cqg{10} (Aug. 1993) 1471--76

\item
R. Loll.
Yang-Mills theory without Mandelstam constraints.
\np{B400} (1993) 126--44

\item
J. Louko.
Holomorphic quantum mechanics with a quadratic Hamiltonian constraint.
gr-qc/9305003, \pr{D48} (Sept. 1993) 2708--27

\item
J. Maluf.
Degenerate triads and reality conditions in canonical gravity.
\cqg{10} (April 1993) 805--9

\item
N. Manojlovi{\' c} and G.A. Mena Marug{\' a}n.
Nonperturbative canonical quantization of nimisuperspace models:
Bianchi types I and II.
gr-qc/9304041, \pr{D48} (Oct. 1993) 3704--19

\item
N. Manojlovi{\' c} and A. Mikovi{\' c}.
Canonical analysis of Bianchi models in the Ashtekar formulation.
\cqg{10} (March 1993) 559--74

\item
D.M. Marolf.
Loop representations for $2+1$ gravity on a torus.
\cqg{10} (Dec. 1993) 2625--47

\item
D. Marolf.
An illustration of 2+1 gravity loop transform troubles.
gr-qc/9305015, {\em Canadian Gen. Rel.} (1993) 256.

\item
H.-J. Matschull.
Solutions to the Wheeler-DeWitt constraint of canonical gravity
coupled to scalar matter fields.
gr-qc/9305025, \cqg{10}:L149-L154, (1993).

\item
H. Nicolai and H.J. Matschull.
Aspects of Canonical Gravity  and Supergravity.
\jgp{11}:15-62, (1993).

\item
O. Obreg{\'o}n, J. Pullin, M.P. Ryan.
Bianchi cosmologies:  New variables and a hidden supersymmetry.
gr-qc/9308001, \pr{D48} (Dec. 1993) 5642--47

\item 
Peter Peld\'an.
Unification of gravity and Yang-Mills theory in 2+1 dimensions.
\np{B395} (1993) 239--62

\item
A. Rendall.
Comment on a paper of Ashtekar and Isham.
\cqg{10} (March 1993) 605--8

\item
A. Rendall.
Unique determination of an inner-product by adjointness
relations in the algebra of quantum observables.
\cqg{10} (Nov. 1993) 2261--69

\item
J. Romano.
Geometrodynamics vs. connection dynamics.
gr-qc/9303032, \grg{25} (Aug.\ 1993) 759--854

\item
J. Romano.
Constraint algebra of degenerate relativity.
gr-qc/9306034, \pr{D48} (Dec. 1993) 5676--83

\item
C. Rovelli.
Area is length of Ashtekar's triad field.
\pr{D47} (Feb.\ 1993) 1703--5

\item
C. Rovelli.
Basis of the Ponzano-Regge-Turaev-Viro-Ooguri quantum-gravity model
is the loop representation basis.
\pr{D48} (Sept. 1993) 2702--07

\item
C. Rovelli.
A generally covariant quantum field theory and a
prediction on quantum measurements of geometry.
\np{B405} (Sept. 1993) 797--815

\item
T. Sano and J. Shiraishi.
The non-perturbative canonical quantization of the $N=1$ supergravity.
\np{B410} (Dec. 1993) 423--47

\item 
L. Smolin.
What can we learn from the study of non-perturbative quantum general
relativity? gr-qc/9211019, in {\em General Relativity and Cosmology 1992},
R.J Gleiser, C Kozameh, O.M. Moreshi eds. (IOP Publishing, 1993).

\item
L. Smolin.
Time, measurement and information loss in quantum cosmology.
in {\em Directions in General Relativity,
Proceedings, Simposium, College Park, USA,
May 1993} B.L. Hu and T. Jacobson (eds),Cambridge U. Press, 1993.
gr-qc/9301016,

\item
R.S. Tate.
Constrained systems and quantization.
In {\em Quantum gravity, gravitational radiation and large scale
structure in the universe}, eds. B.R.\ Iyer, S.V. Dhurandhar and K.
Babu Joseph (1993)

\item
T. Thiemann.
On the solution of the initial value constraints for
general relativity coupled to matter in terms of Ashtekar's variables.
\cqg{10} (Sept. 1993) 1907--21

\item
T. Thiemann and H.A. Kastrup.
Canonical quantization of spherically symmetric
gravity in Ashtekar's self-dual representation.
\np{B399} (June 1993) 211--58

\item
D.A. Ugon, R. Gambini and P. Mora.
Link invariants for intersecting loops.
\pl{B305} (May 1993) 214--22

\item
J. Zegwaard.
Physical interpretation of the loop representation for non-perturbative
quantum gravity.
\cqg{10} suppl. (Dec. 1993) S273--6

\item
J. Zegwaard.
The weaving of curved geometries.
\pl{B300} (Feb. 1993) 217--222

\newpage

\section*{1994}

\item
D. Armand-Ugon, R. Gambini, J. Griego, and L. Setaro.
Classical loop actions of gauge theories,
hep-th/9307179, \pr{D50} (1994) 5352

\item
A. Ashtekar.
Overview and outlook.
CGPG-94/1-1, gr-qc/9403038, in  J Ehlers and H. Friedrich, eds,
{\em Canonical Gravity: From Classical to
Quantum}. Springer-Verlag Berlin (1994).

\item
A. Ashtekar and J. Lee.
Weak field limit of General Relativity in terms of new variables:
A Hamiltonian framework.
CGPG-94/8-3, \ijmp{D3}: 675-694, (1994).

\item
A. Ashtekar and J. Lewandowski.
Representation theory of analytic holonomy $C^{*}$-algebras.
In {\em Knots and Quantum Gravity}, ed. J. Baez, Oxford U. Press, 1994.
gr-qc/9311010.

\item
A. Ashtekar and R. Loll.
New loop representations for $2+1$ gravity.
CGPG-94/5-1, gr-qc/9405031, \cqg{11}: 2417-2434, 1994

\item
A. Ashtekar, D. Marolf and J. Mour\~ao.
Integration on the space of connections modulo gauge transformations.
{\em Proc. Lanczos International Centenary Conference}. 
J. Brown et al (eds), SIAM, Philadelphia, 1994.
CGPG-94/3-4, gr-qc/9403042.

\item
A. Ashtekar and R.S. Tate.
An algebraic extension of Dirac quantization:  Examples.
CGPG-94/6-1, gr-qc/9405073. \jmp{35} (1994), 6434

\item
A. Ashtekar and M. Varadarajan.
A striking property of the gravitational Hamiltonian.
CGPG-94/8-3, gr-qc/9406040, \pr{D52} (1994), 4944

\item
J. Baez.
Generalized Measures in Gauge Theory.
{\em Lett. Math. Phys.} {\bf 31} (1994) 213--223

\item
J.C. Baez.
Strings, loops, knots, and gauge fields. in {\em Knots and Quantum Gravity,
Proceedings, Workshop, Riverside, 1993}.
J.C. Baez (ed), Clarendon, 
Oxford U.K. (1994).
hep-th/9309067.

\item
J.C. Baez.
Diffeomorphism-invariant generalized measures on the space of connections
modulo gauge transformations.
in {\em The Proceedings of the Conference on Quantum Topology},
L. Crane and D. Yetter (eds) World Scientific, Singapore, 1994.
hep-th/9305045.

\item
J.F. Barbero G.
Real-polynomial formulation of general relativity in terms of connections.
\pr{D49} (June 1994) 6935--38

\item
J.F. Barbero.
General Relativity as a Theory of 2 Connections.
CGPG-93/9-5, gr-qc/9310009, \ijmp{D3} (1994) 379--392

\item
J.F. Barbero G. and M. Varadarajan.
The phase space of $(2+1)$-dimensional gravity in the Ashtekar formulation.
\np{B415} (Mar. 1994) 515--530,
gr-qc/9307006.

\item
C. Di Bartolo, R. Gambini, J. Griego and J. Pullin.
Extended loops:  A new arena for nonperturbative quantum gravity.
\prl{72} (June 1994) 3638--41

\item
Y. Bi and J. Gegenberg
Loop variables in topological gravity.
gr-qc/9307031, \cqg{11} (Apr. 1994) 883--96

\item
R. Borissov.
Weave states for plane gravitational waves.
\pr{D49} (Jan. 1994) 923--29

\item
O. B\"ostrom. Loop variables and degeneracy. in
{\em Proceedings of the VII J.A. Swieca Summer School on Particles and
Fields, Sao Paolo, Brazil, 1993}, World Scientific (1994).

\item
B. Br\"ugmann.
Loop Representations.
MPI-Ph/93-94, gr-qc/9312001.In  J Ehlers and H. Friedrich, eds,
{\em Canonical Gravity: From Classical to
Quantum}. Springer-Verlag Berlin (1994).

\item
R. Capovilla and J. Guven.
Super-Minisuperspace and New Variables.
CIEA-GR-9401, gr-qc/9402025,   \cqg{11} (1994) 1961--70

\item
R. Capovilla and O. Obregon
No quantum Superminisuperspace with $\Lambda\neq 0$. \pr{D49} (1994),
6562

\item
S. Carlip.
Geometrical structures and loop variables in $2+1)$-dimensional gravity.
In {\em Knots and Quantum Gravity}, ed. J. Baez, Oxford U. Press, 1994.
UCD-93-30, gr-qc/9309020.

\item
S.M. Carroll and G.B. Field.
Consequences of propogating torsion in connection-dynamic 
theories of gravity.
\pr{D50}: 3867-3873, 1994.
gr-qc/9403058.

\item
S. Chakraborty and P. Peld\'an.
Towards a unification of gravity and Yang-Mills theory.
CGPG-94/1-3, gr-qc/9401028, \prl{73} (1994) 1195.

\item
S. Chakraborty and P. Peld\'an.
Gravity and Yang-Mills theory: two faces of the same theory?
CGPG-94/2-2, gr-qc/9403002,
{\em Int. J. of Mod. Phys.}{\bf D}3 (1994) 695.

\item
L.N. Chang and C. Soo. Superspace Dynamics and perturbations around
`emptiness'. \ijmp{D3}:529 (1994).

\item
L. Crane.
Topological field theory as the key to quantum gravity.
In {\em Knots and Quantum Gravity}, ed. J. Baez, Oxford U. Press, 1994.

\item
G. Esposito and H.A. Morales-T\'ecotl. Self Dual Action for Fermionic
Fields and Gravitation. gr-qc/9506073, {\em Nuov. Cimiento}
{\bf 109B}: 973-982 (1994).

\item
G. Esposito, H.A. Morales-T\'ecotl and G. Pollifrone. Boundary Terms for
Massless Fermionic Fields. gr-qc/9506075, {\em Found. Phys. Lett.} {\bf 7}:
303-308 (1994).

\item
S. Frittelli, S. Koshti, E.T. Newman and C. Rovelli.
Classical and quantum dynamics of the Faraday lines of force.
\pr{D49} (June 1994) 6883--91

\item
G. F{\" u}l{\" o}p.
About a super-Ashtekar-Renteln ansatz.
gr-qc/9305001, \cqg{11} (Jan. 1994) 1--10

\item
R. Gambini and J. Pullin.
The Gauss linking number in quantum gravity.
In {\em Knots and Quantum Gravity}, ed. J. Baez, Oxford U. Press, 1994.

\item
D. Giulini.
Ashtekar variables in classical general relativity.
THEP-93/31, gr-qc/9312032. In  J Ehlers and H. Friedrich, eds,
{\em Canonical Gravity: From Classical to
Quantum}. Springer-Verlag Berlin (1994).

\item
J.N. Goldberg and D.C. Robinson.
Linearized constraints in the connection representation:
Hamilton-Jacobi solution.
\pr{D50}: 6338-6343, 1994.
gr-qc/9405030.

\item
S. Hacyan.
Hamiltonian Formulation of General Relativity
in terms of Dirac Spinors.
\grg{26} (Jan. 1994) 85--96

\item
J. Hallin.
Representations of the SU(N) T-Algebra and the Loop Representation in
1+1 Dimensions.  \cqg{11} 1615--1629

\item
H.-L. Hu.
$W_{1+\infty}$, KP and loop representation of four dimensional gravity.
\pl{B324} (Apr. 1994) 293--98

\item
V. Husain.
Self-dual gravity and the chiral model.
\prl{72} (Feb. 1994) 800--03

\item
V. Husain.
Self-dual gravity as a two-dimensional theory and conservation laws.
\cqg{11} (Apr. 1994) 927--38

\item
V. Husain.
Observables for spacetimes with two Killing field symmetries.
Alberta-Thy-55.93, gr-qc/9402019, \pr{D50}, (1994), 6207

\item
G. Immirzi.
Regge Calculus and Ashtekar Variables.
DFUPG 83/94, gr-qc/9402004, \cqg{11} (1994) 1971--79

\item
J. Iwasaki and C. Rovelli.
Gravitons from Loops -- Nonperturbative Loop-Space Quantum-Gravity
Contains the Graviton-Physics Approximation.  \cqg{11} (1994) 1653--76

\item
H.A. Kastrup and T. Thiemann.
Spherically symmetric gravity as a completely integrable system.
PITHA 93-35, gr-qc/9401032, \np{B425}: 665-685, (1994).

\item
L.H. Kauffmann.
Vassiliev invariants and the loop states in quantum gravity.
In {\em Knots and Quantum Gravity}, ed. J. Baez, Oxford U. Press, 1994.

\item
J. Lewandowski.
Topological Measure and Graph-Differential Geometry on the
Quotient Space of Connections.
gr-qc/9406025, \ijmp{D3} (1994) 207--210

\item
J. Lewandowski.
Differential geometry for the space of connections modulo gauge
transformations. In {\em the proceedings of  Cornelius Lanczos
International Centenary Conference}. J. Brown, et al (eds), SIAM
Phyladelphya, 1994.
gr-qc/9406026

\item
R. Loll. 
The loop formulation of gauge theory and gravity.
In {\em Knots and Quantum Gravity}, ed. J. Baez, Oxford U. Press, 1994.

\item
R. Loll.
Gauge theory and gravity in the loop formulation. 
CGPG-94/1-1. In  J Ehlers and H. Friedrich, eds,
{\em Canonical Gravity: From Classical to
Quantum}. Springer-Verlag Berlin (1994).

\item
J. Louko and D. Marolf.
Solution space of 2+1 gravity on ${\bf R} \times T^2$ in Witten's
connection formulation.
gr-qc/9308018, \cqg{11} (1994), 315

\item
J. M{\" a}kel{\" a}.
Phase space coordinates and the Hamiltonian constraint of Regge calculus.
\pr{D49} (Mar. 1994) 2882--96

\item
H.-J. Matschull.
About loop states in supergravity.
DESY-94-037, gr-qc/9403034, \cqg{11}: 2395-2410, (1994).

\item
H.-J. Matschull and H. Nicolai.
Canonical quantum supergravity in three dimensions.
gr-qc/9306018, \np{B411} (Jan. 1994) 609--46

\item
G.A. Mena Marug{\' a}n.
Reality conditions in non-perturbative quantum cosmology.
\cqg{11} (Mar. 1994) 589--608

\item
G. Modanese.
Wilson Loops in 4-Dimensional Quantum-Gravity.
\pr{D49} (1994) 6534--42

\item
H.A. Morales-T{\' e}cotl and C. Rovelli.
Fermions in quantum gravity.
\prl{72} (June 1994) 3642--45

\item
J.M. Nester, R.-S. Tung and Y.Z. Zhang.
Ashtekar's new variables and positive energy.
\cqg{11} (Mar. 1994) 757--66

\item
J.A. Nieto, O. Obreg\'on and J. Socorro.
The gauge theory of the de-Sitter group and Ashtekar formulation.
IFUG-94-001, gr-qc/9402029, \pr{D50} (1994), 3583

\item
P. Peld{\' a}n.
Actions for gravity, with generalizations:  a review.
gr-qc/9305011, \cqg{11} (May. 1994) 1087--1132

\item 
Peter Peld\'an.
Ashtekar's variables for arbitrary gauge group.
G\"oteborg ITP 92-17, \pr{D46}, R2279.

\item
P. Peld\'an.
Real formulations of complex gravity and a complex 
formulation of real gravity.
GCPG-94/4-6, gr-qc/9405002, \np{B430}, (1994), 460

\item
J. Pullin.
Knot theory and quantum gravity: a primer.
in {\em Fifth Mexican School of Particles and
Fields}. J.L. Lucio and M. Vargas (eds),
AIP Conference Proceedings 317, AIP Press (1994).
hep-th/9301028.

\item
A. Rendall.
Adjointness relations as a criterion for choosing an inner product.
MPA-AR-94-1, gr-qc/9403001.  In  J Ehlers and H. Friedrich, eds,
{\em Canonical Gravity: From Classical to
Quantum}. Springer-Verlag Berlin (1994).

\item
C. Rovelli and L. Smolin.
The physical Hamiltonian in nonperturbative quantum gravity.
\prl{72} (Jan. 1994) 446--49

\item
D.C. Salisbury and L.C. Shepley.
A connection approach to numerical relativity.
\cqg{11}: 2789-2806, 1994.
gr-qc/9403040.

\item
P. Schaller, J. Strobl.
Canonical quantization of two-dimensional gravity with torsion
and the problem of time.
\cqg{11} (Feb. 1994) 331--46

\item
L. Smolin
Finite Diffeomorphism-Invariant Observables in Quantum Gravity.
SU-GP-93/1-1, gr-qc/9302011, \pr{D49} (1994) 4028--40

\item
J. Tavares.
Chen integrals, generalized loops and loop calculus.
\ijmp{A9}: 4511-4548, 1994.

\item
T. Thiemann.
Reduced Phase-Space Quantization of Spherically Symmetrical Einstein-Maxwell
Theory Including a Cosmological Constant.
\ijmp{D3} (1994) 293-298

\newpage

\section*{1995}

\item
D. Armand-Ugon, R. Gambini and P. Mora.
Intersecting braids and intersecting knot theory.
{\em Journal of Knot theory and its ramifications}
{\bf 4}:1, 1995.
hep-th/9309136.

\item A. Ashtekar.
Mathematical problems of non-perturbative quantum general relativity.
Published in: {\em Gravitation and Quantization (Les Houches,
Session LVIII, 1992}. Ed. B. Julia (Elsevier, Amsterdam, 1995)

\item
A. Ashtekar.
Recent Mathematical Developments in Quantum General Relativity.
In {\em The Proceedings of PASCOS-94}. K.C. Wali (ed), 
World Scientific, 1995.
gr-qc/9411055.

\item 
A. Ashtekar and J. Lewandowski. 
Differential geometry on the space
of connections via graphs and projective limits.
CGPG-94/12-4, hep-th/9412073, {\em J. Geom. Phys.} {\bf 17} (1995), 191

\item
A. Ashtekar and J. Lewandowski.
Projective techniques and functional integration.
CGPG/94/10-6, gr-qc/9411046, \jmp{36} (1995), 2170

\item
A. Ashtekar, J. Lewandowski, D. Marolf, J. Mour\~ao and T. Thiemann.
Quantization of diffeomorphism invariant theories of connections 
with local degrees of freedom.
gr-qc/9504018, \jmp{36} (1995), 519

\item
G. Au.
The quest for quantum gravity.
{\em Current Science}, {\bf 69}: 499-518, 1995.
gr-qc/9506001.

\item
J.C. Baez.
Link invariants, holonomy algebras and functional integration.
{\em Jour. Funct. Analysis} {\bf 127}:108, 1995.

\item
J. C. Baez, J. P. Muniain, and Dardo D. Piriz.
Quantum gravity Hamiltonian for manifolds with boundary.
gr-qc/9501016, \pr{D52} (1995), 6840

\item
J. Fernando Barbero G.
Solving the constraints of general relativity.\cqg{12}:L5-L10, 1995.
 CGPG-94/11-3, gr-qc/9411016.

\item
J. Fernando Barbero G. 
Reality Conditions and Ashtekar Variables: a Different Perspective,
\pr{D51}:5498-5506, 1995.  gr-qc/9410013.

\item
J. Fernando Barbero G. 
Real Ashtekar Variables for Lorentzian Signature Space-times. \pr{D51}:5507
-5510 (1995).  gr-qc/9410014.

\item
J. F. Barbero and M. Varadarajan. 
Homogeneous (2+1)-Dimensional Gravity in the Ashtekar Formulation,
\np{B456}:355-376, 1995. gr-qc/9507044.

\item
C. Di Bartolo, R. Gambini, and J. Griego.
Extended loop representation of quantum gravity, gr-qc/9406039,
\pr{D51} (Jan. 1995) 502

\item
C. Di Bartolo, R. Gambini, J. Griego and J. Pullin.
The space of states of quantum gravity in terms of loops and extended
loops:  some remarks.
\jmp{36}: 6511, 1995.
 gr-qc/9503059.

\item
P.A. Blaga, O. Moritsch and H. Zerrouki.
Algebraic structure of gravity in Ashtekar's variables.
TUW 94-15, hep-th/9409046, \pr{D51} (Mar. 1995) 2792

\item
B. Br\"ugmann.
On a geometric derivation of Witten's identity for Chern-Simons
theory via loop deformations.
{\em Int. J. Theor. Phys.} {\bf 34}:145, 1995.
hep-th/9401055.

\item
G. Esposito and C. Stornaiolo.
Space-time covariant form of 
Ashtekar's constraints. 
\ncim{110B}:1137-1152, 1995.
gr-qc/9506008.

\item
K. Ezawa.
Combinatorial solutions to the Hamiltonian constraint in 
(2+1)-dimensional Ashtekar gravity.
gr-qc/9506043, \np{B459}: 355-392 (1995).

\item
T.J. Foxon.
Spin networks, Turaev - Viro theory and the loop representation.
\cqg{12} (April 1995) 951

\item
R. Gambini, A. Garat and J. Pullin.
The constraint algebra of quantum gravity in the loop representation.
CGPG-94/4-3, gr-qc/9404059, {\em Int. J. Mod. Phys.} {\bf D4} (1995) 589

\item
H. Garcia-Compean and T. Matos. 
Solutions in Self-dual Gravity constructed Via Chiral Equations. 
\pr{D 52}: 4425-4429, 1995.
 hep-th/9409135.

\item
J.N. Goldberg and C. Soteriou.
Canonical General Relativity on a null surface with coordinate and
gauge fixing.
\cqg{12}: 2779-2798, 1995.
gr-qc/9504043.

\item
G. Gonzalez and R. Tate.
Classical analysis of Bianchi types I and II in Ashtekar variables.
\cqg{12}:1287-1304, 1995.
 gr-qc/9412015.

\item
I. Grigentch and D.V. Vassilevich.
Reduced phase space quantization of Ashtekar's
gravity on de Sitter background.
\ijmp{D4}: 581-588, 1995.
gr-qc/9405023.

\item
V. Husain. 
The affine symmetry of self-dual gravity. 
\jmp{36}: 6897, 1995.
 hep-th/9410072.
 
\item
R.A. d'Inverno and J.A. Vickers.
2+2 decomposition of Ashtekar variables.  \cqg{12} (Mar. 1995) 753

\item
J. Iwasaki and C. Rovelli.
Gravitons as embroidery on the weave.
{\em Int. J. Mod. Phys}{\bf D1} (1995) 533

\item
R. Loll.
Independent loop invariants for $2+1$ gravity.
CGPG-94/7-1, gr-qc/9408007, \cqg{12}:1655-1662, (1995).

\item
R. Loll.
Quantum Aspects of $2+1$ Gravity. gr-qc/9503051, \jmp{36}:6494-6509, (1995).

\item
R. Loll.
Non-perturbative solutions for lattice quantum gravity.
gr-qc/9502006, \np{B444}:614-640, (1995).

\item
R. Loll.
The volume operator in discretized quantum gravity.
gr-qc/9506014, \prl{75}:3048-3051, (1995).

\item
R. Loll, J.M. Mour\~ao and J.N. Tavares.
Generalized coordinates on the phase space of Yang-Mills theory.
CGPG-94/4-2, gr-qc/9404060, \cqg{12}:1191-1198, (1995).

\item
J. Louko.
Chern-Simons functional and the no-boundary proposal in Bianchi IX
quantum cosmology.
\pr{D51}: 586-590, 1995.
gr-qc/9408033.

\item
S. Major and L. Smolin.
Cosmological histories for the new variables.
CGPG-94/2-1, gr-qc/9402018, \pr{D51} (1995), 5475

\item
D. Marolf and J.M. Mour\~ao.
On the support of the Ashtekar-Lewandowski measure.
CGPG-94/3-1, hep-th/ 9403112, {\em Comm. Math. Phys.}{\bf 170} (1995),
583

\item
H.J. Matschull.
Tree-dimansional Canonical Quantum Gravity. gr-qc/9506069, \cqg{12}: 2621-2704,
(1995).

\item
H.J. Matschull.
New Representation and a Vacuun State for Canonical Quantum Gravity.
gr-qc/9412020, \cqg{12}: 651-676, (1995).

\item
G.A. Mena Marug\'an.
Is the exponential of the Chern-Simons action a normalizable physical state?
CGPG-94/2-2, gr-qc/9402034, \cqg{12} (Feb. 1995) 435

\item
S. Mizoguchi.
The Geroch group in the Ashtekar formulation.
\pr{D51}: 6788-6802, 1995.
gr-qc/9411018.

\item
H.A. Morales-T\'ecotl and C. Rovelli. Loop Space Representation of Quantum
Fermions and Gravity. \np{B451} (1995) 325. 

\item
R. de Pietri and C. Rovelli. Eigenvalues of the Weyl Operator as
Observables of General Relativity. \cqg{12}: 1279-1286 (1995).

\item
J. Pullin.
Recent Developments in Canonical Quantum Gravity.
CAM-94 Physics Meeting, in AIP Conf. Proc {\bf 342}, 459, ed.  Zepeda A
 (AIP Press, Woodbury, New York), 1995.

\item
M. Reisenberger.
New Constraints for Canonical General Relativity.
\np{B457}: 643-687, 1995.
gr-qc/9505044.

\item
C. Rovelli.
Outline of a generally covariant quantum field theory and a quantum 
theory of gravity.
gr-qc/9503067, \jmp{36}: 6529-6547 (1995).

\item
C. Rovelli and L. Smolin. Spin Networks and Quantum Gravity. gr-qc/9505006,
\pr{D52}: 5743-5759.

\item
C. Rovelli and L. Smolin.
Discreteness of area and volume in quantum gravity.
\np{B442}: 593-622 (1995). Erratum: \np{B456}:734 (1995).
gr-qc/9411005,

\item
L. Smolin. Linking TQFT and Nonperturbative Quantum Gravity. gr-qc/9505028,
\jmp{36}: 6417-6455 (1995).

\item
L. Smolin and C. Soo.
The Chern-Simons invariant as the natural time variable for
classical and quantum cosmology.
CGPG-94/4-1, gr-qc/9405015, \np{B449}: 289-316 (1995). 

\item
C. Soo.
Self-dual variables, positive semi-definite action, and discrete
transformations in 4-d quantum gravity.
gr-qc/9504042, \pr{D52}: 3484-3493 (1995).

\item
I. A. B. Strachan. 
The symmetry structure of the anti-self-dual Einstein 
hierarchy.
\jmp{36}: 3566-3573, 1995.
 hep-th/9410047.

\item
T. Thiemann.
Generalized boundary conditions for general relativity for
the asymptotically flat case in terms of Ashtekar's variables.
\cqg{12} (Jan. 1995) 181

\item
T. Thiemann.
Complete quantization of a diffeomorphism invariant field theory.
\cqg{12} (Jan. 1995) 59

\item
T. Thiemann
The reduced phase space of spherically symmetric
Einstein-Maxwell theory including a cosmological constant.
\np{B436} (Feb. 1995) 681  

\item
T. Thiemann.
An account of transforms on $\overline{{\cal A}/{\cal G}}$.
{\em Acta Cosmologica} {\bf 21}: 145-167, 1995.
gr-qc/9511050.

\item
R.S. Tung and T. Jacobson.
Spinor one forms as gravitational potentials
\cqg{12}: L51-L55, 1995.
gr-qc/9502037.

\newpage
\section*{1996}

\item
D. Armand-Ugon, R. Gambini, O. Obregon and J. Pullin.
Towards a loop representation for quantum canonical supergravity.
hep-th/9508036, \np{B460} (1996), 615

\item
J. M. Aroca, H. Fort and R. Gambini
The Path Integral for The Loop Representation of Lattice Gauge Theories.
\pr{D54}:7751-7756, 1996. hep-th/9605068

\item
A. Ashtekar.
A Generalized Wick Transform for Gravity.
gr-qc/9511083, \pr{D53}: 2865-2869, (1996),

\item
A. Ashtekar, J. Lewandowski, D. Marolf, J. Mour\~ao and T. Thiemann.
A manifestly gauge-invariant approach to quantum theories
of gauge fields.
in {\em Geometry of Constrained Dynamical
Systems. Proceedings, Conference, Cambridge, UK, 1994}. 
J.M. Charap (ed), Cambridge University Press, 1996. 
hep-th/9408108

\item
A. Ashtekar, J. Lewandowski, D. Marolf, J. Mour\~ao and T. Thiemann.
Coherent State Transforms for Spaces of Connections.
{\em J. Functional Analysis} {\bf 135}: 519-551, 1996.
 gr-qc/9412014.

\item
J. C. Baez.
Spin networks in gauge theory.
{\em Advances in Mathematics} {\bf 117}: 253, 1996.
gr-qc/9411007.

\item
J. C. Baez.
Spin networks in nonperturbative quantum gravity. In {\em Interface
of Knot Theory and Physics}, L. Kauffman (ed), American Mathematical
Society, Providence, Rhode Island, 1996.
gr-qc/9504036.

\item
J. Fernando Barbero G.
From Euclidean to Lorenzian General Relativity: The Real Way, 
\pr{D54}:1492-1499, 1996. gr-qc/9605066

\item
J. F. Barbero and M. P. Ryan.
Minisuperspace Examples of Quantization Using Canonical Variables of 
the Ashtekar Type: Structure and Solutions. \pr{D53}:5670-5681, 1996.
gr-qc/9510030.

\item
M. Barreira, M. Carfora and C. Rovelli.
Physics with nonperturbative quantum gravity: radiation from a 
quantum black hole. \grg{28}:1293-1299, 1996.
gr-qc/9603064.

\item
R. Borissov, S. Major and L. Smolin,
The Geometry of Quantum Spin Networks.
\cqg{13}: 3183-3196, 1996.
gr-qc/9512043.

\item
L. N. Chang and Chopin Soo.
Chiral fermions, gravity and GUTs.
CGPG-94/9-3, hep-th/9411064, \pr{D53}: 5682-5691
(1996).

\item
R. De Pietri and C. Rovelli.
Geometry Eigenvalues and Scalar Product from Recoupling Theory in Loop 
Quantum Gravity.
gr-qc/9602023, \pr{D54}: 2664-2690 (1996).

\item
K. Ezawa. A Semiclassical Interpretation of the Topological Solutions
for Canonical Quantum Gravity. gr-qc/9512017, \pr{D53}: 5651-5663 (1996).

\item
K. Ezawa.
Multi-plaquette solutions for discretized Ashtekar gravity.
gr-qc/9510019, {\em Mod. Phys. Lett.} {\bf A}: 349-356 (1996).

\item
K. Ezawa.
Ashtekar's formulation for $N=1,2$ supergravities as ``constrained" BF
theories.
hep-th/9511047, {\em Prog. Theor. Phys.} {\bf 95}: 863 (1996).

\item
S. Fritelli, L. Lehner and C. Rovelli. The complete spectrum of the area
from recoupling theory in loop quantum gravity. 
\cqg{13}:2921-2932,1996.  gr-qc/9608043.

\item
R. Gambini and J. Pullin.
A rigorous solution of the quantum Einstein equations.
\pr{D54}:5935-5938, 1996.
gr-qc/9511042.

\item
R. Gambini and J. Pullin.
Knot theory and the dynamics of quantum general
relativity.
\cqg{13}:L125, 1996.
gr-qc/9603019.

\item
H. Garcia-Compean, J. Plebanski and M. Przanowski.
From Principal chiral model to selfdual gravity.
\mpl{A11}: 663-674, 1996.

\item
J.N. Goldberg.
Generalized Hamilton-Jacobi transformations: gauge and
diffeomorphism constraint.
\pr{D54}: 4997-5001, 1996.

\item
J. Griego.
Extended knots and the space of states of quantum gravity.
\np{B473}:291-307, 1996.
gr-qc/9601007.

\item
J. Griego.
The Kauffman Bracket and the Jones Polynomial
in Quantum Gravity. \np{B467}:332-354, 1996.
gr-qc/9510050.

\item
J. Griego.
Is the Third Coefficient of the Jones Knot Polynomial
a Quantum State of Gravity?
\pr{D53}:6966-6978, 1996.
gr-qc/9510051.

\item
N. Grot and C. Rovelli.
Moduli-space of knots with intersections.
gr-qc/960410, \jmp{37}: 3014-3021 (1996).

\item
H. Ishihara, H. Kubotani and T. Fukuyama.
Gravitational Instantons in Ashtekar's Formalism.
\ijmp{A11}: 2707-2720, 1996.
gr-qc/9509009.

\item
T. Jacobson.
1+1 sector of 3+1 gravity.
\cqg{13}:L111-L116, 1996.
erratum-ibid {\bf 13}:3269, 1996.
gr-qc/9604003.

\item
S. Holst.
Barbero's Hamiltonian derived from a generalized Hilbert-Palatini action.
\pr{D53}: 5966-5969, 1996.
gr-qc/9511026.

\item
V. Husain.
Einstein's equations and the chiral model.
gr-qc/9602050, \pr{D53} (1996), 4327

\item
V. Husain.
General Covariance, and Supersymmetry Without
Supersymmetry. \pr{D54}:7849-7856, 1996.
hep-th/9609009.

\item
G. Immirzi
Quantizing Regge Calculus.
\cqg{13}: 2385-2394, 1996.
gr-qc/9512040.

\item
K. Krasnov.
Quantum loop representation for fermions coupled to
Einstein-Maxwell field.
\pr{D53}: 1874-1888, 1996.

\item
S. R. Lau.
New Variables, the gravitational action, and boosted quasilocal 
stress-energy-momentum.
\cqg{13}: 1509-1540, 1996.
gr-qc/9504026.

\item
L. Leal.
Electric-Magnetic duality and the `Loop Representation'
in Abelian gauge theories.
\mpl{A11}: 1107-1114, 1996.

\item
R. Loll.
Spectrum of the volume operator in quantum gravity.
gr-qc/9511030 \np{B460}:143-154, (1996).

\item
R. Loll.
A real alternative to quantum gravity in loop space.
\pr{D54}: 5381, 1996.
gr-qc/9602041. 

\item
S. Major and L. Smolin.
Quantum Deformation of Quantum Gravity.
\np{B473}: 267-290, 1996.
gr-qc/9512020.

\item
H.J. Matschull.
Causal Structure and Diffeomorphisms in Ashteker's Gravity.
gr-qc/9511034, \cqg{13}: 765-782, (1996).

\item
G. Mena Marug\'an.
Involutions on the algebra of physical observables from
reality conditions.
\jmp{37}: 196-205, 1996.
gr-qc/9506038.

\item
H.A. Morales-T\'ecotl, L.F. Urrutia and J.D. Vergara. Reality
Conditions for Ashtekar Variables as Dirac Constraints. 
\cqg{13}: 2933-2940, 1996.
gr-qc/9607044.

\item
P. Peldan.
Large Diffeomorphisms in (2+1)-Quantum Gravity on the Torus.
CGPG-95/1-1, gr-qc/9501020 \pr{D53} (1996), 3147

\item
C. Rovelli. Loop Quantum Gravity and Black hole Physics.
{\em Helv. Phys. Acta} {\bf 69}:582-611, 1996.
gr-qc/9608032.

\item
C. Rovelli.
Black hole entropy from loop quantum gravity.
\prl{77}:3288-3291, 1996.
gr-qc/9603063.

\item
T. Thiemann.
Reality conditions inducing transforms for quantum gauge 
field theory and quantum gravity.
\cqg{13}: 1383-1404, 1996.
gr-qc/9511057.

\item
T. Thiemann.
Anomaly-Free Formulation of Nonperturbative
Four-dimensional Lorentzian Quantum Gravity.
\pl{B380}: 257-264, 1996.
gr-qc/9606088.

\item
H. Waelbroeck and J.A. Zapata.
2+1 covariant lattice theory and `t Hooft's
formulation.
\cqg{13}: 1761-1768, 1996.
gr-qc/9601011.

\item
G. Yoneda and H. Shinkai.
Constraints and reality conditions in the Ashtekar formulation of general 
relativity.
\cqg{13}: 783-790, 1996.
gr-qc/9602026. 

\item
J. A. Zapata.
Topological lattice gravity using self dual
variables. \cqg{13}: 2617-2634, 1996.
gr-qc/9603030.

\newpage
\section*{1997}

\item
A. Ashtekar and A. Corichi.
Photon Inner Product and the Gauss Linking
Number. \cqg{14}:A43-A53, 1997. gr-qc/9608017.

\item
A. Ashtekar and J. Lewandowski.
Quantum Theory of Geometry I: Area Operators. \cqg{14}:A55-A81, 1997.
gr-qc/9602046. 

\item
R. Borissov.
Regularization of the Hamiltonian constraint and the closure 
of the constraint algebra.
\pr{D55}:2059-2068, 1997.
gr-qc/9411038.

\item
R. De Pietri.
On the relation between the connection and the loop representation of 
quantum gravity. \cqg{14}:53-69, 1997.
gr-qc/9605064.

\item
B.P. Dolan and K.P. Haugh.
A Covariant Approach to Ashtekar's Canonical Gravity, 
\cqg{14}:477-488, 1997.

\item
K. Krasnov. Geometrical entropy from loop quantum gravity.
\pr{D55}:3505, 1997.
(Title changed in journal. Other title: Counting Surface States 
in the loop quantum gravity.
Also known as: The Bekenstein bound and nonperturbative
quantum gravity.)
gr-qc/9603025.

\item
J. Lewandowski.
Volume and Quantizations. \cqg{14}:71-76, 1997.
gr-qc/9602035.

\item
D. E. Neville.
Open-flux solutions to the constraints for plane gravity waves.
\pr{D55}: 766-780, 1997.
gr-qc/9511061.

\item
D. E. Neville.
Closed flux solutions to the quantum constraints for
plane gravity waves.
\pr{D55}: 2069-2075, 1997.
gr-qc/9607053.

\newpage
\section*{Preprints older than 12 months}

\item
J. Baez. 
Knots and Quantum Gravity: Progress and Prospects. 
To appear in {\em Proceedings of the VIIth Marcel Grossman 
Meeting}.
gr-qc/9410018.

\item
I. Bengtsson.
Form connections.
gr-qc/9305004

\item
I. Bengtsson.
Ashtekar's variables. 
Goteborg-88-46 preprint (November 1988).

\item
I. Bengtsson.
Curvature tensors in an exact solution of Capovilla's equations.
Goteborg-91-5 (February 1991).

\item
I. Bengtsson.
Ashtekar's variables and the cosmological constant.
Goteborg preprint, 1991.

\item
I. Bengtsson.
Form Geometry and the `t Hooft-Plebanski
action.
gr-qc/950210

\item
Ola Bostr\"om.
Some new results about the cosmological constants.
G\"oteborg preprint ITP91-34

\item
O. Bostr\"om, M. Miller and L. Smolin.
A new discretization of classical and quantum general relativity.
G\"oteborg ITP 94-5, SU-GP-93-4-1, CGPG-94/3-3, gr-qc/9304005.

\item
R. Brooks.
Diff($\Sigma$) and metrics from Hamiltonian-TQFT.
MIT preprint CTPH2175

\item
Greorgy Burnett, Joseph D. Romano, and Ranjeet S. Tate.
 Polynomial coupling of matter to gravity using {A}shtekar variables.
Syracuse preprint.

\item
R. Capovilla, J. Dell and T. Jacobson.
The initial value problem in light of Ashtekar's variables.
UMDGR93-140, gr-qc/9302020

\item
Steven Carlip.
 2+1 dimensional quantum gravity and the Braid group.
 Talk given at the Workshop on Physics, Braids and Links, Banff Summer
  School in Theoretical Physics, August 1989.

\item
L. Chang and C. Soo.
The standard model with gravity couplings.
CGPG-94/6-2.

\item
L. Chang and C. Soo.
Einstein manifolds in Ashtekar variables: explicit examples.
hep-th/9207056

\item
L. Crane.
Categorical physics.
Preprint ???.

\item
Ch. Devchand and V. Ogievetsky.  
Self-dual Gravity Revisited. JINR-E2-94-342, hep-th/9409160.

\item
G. Esposito.
 Mathematical structures of space-time.
 Cambridge preprint DAMTP-R-gols, to appear in Fortschritte der Physik.

\item
R. Floreanini, R. Percacci and E. Spallucci.
Why is the metric non-degenerate?
SISSA 132/90/EP preprint (October 1990).

\item
R. Gambini, and L. Leal.
Loop space coordinates, linear representations of the diffeomorphism group 
and knot invariants.

\item
Sucheta Koshti and Naresh Dadhich.  
The General Self-dual solution of the Einstein Equations.  IUCAA 94/29, 
gr-qc/9409046.

\item
B.~Grossmann.
 General relativity and a non-topological phase of topological
  {Y}ang-{M}ills theory.
 Inst. for Advanced Studies, Princeton, 1990 preprint.

\item
G. Harnett.
Metrics and dual operators.
Florida Atlantic University preprint, 1991.

\item
G. Horowitz.
Ashtekar's approach to quantum gravity. 
University of California preprint, 1991.

\item
W. Kalau.
Ashtekar formalism with real variables. 
U. Of Wuppertal NIKHEF-H/91-03 Amsterdam preprint (December 1990).

\item
K. Kamimura and T. Fukuyama.
Massive analogue of Ashtekar-CDJ action.
gr-qc/9208010

\item
V. Khatsymovsky.
On polynomial variables in general relativity.
BINP 93-41, gr-qc/9310005.

\item
A. Kheyfets and W. A. Miller.
E. Cartan's moment of rotation in Ashtekar's theory of gravity.
Los Alamos preprint LA-UR-91-2605 (1991).

\item
S. Koshti and N. Dadhich.
 Gravitational instantons with matter sources using
Ashtekar variables.
 Inter Univ. Centre for Astron. and Astrophysics, Pune, India.
June 1990 preprint.

\item
C. Kozameh, W. Lamberti, and E.T. Newman
Holonomy and the Einstein equations.
???

\item
R. Loll.
Chromodynamics and gravity as theories on loop space.
CGPG-93/9-1, hep-th/9309056.

\item
R. Loll.
An example of loop quantization.
CGPG-94/7-2.

\item
R. Loll
Wilson loop coordinates for $2+1$ gravity.
CGPG-94/8-1.

\item
R. Loll, J. Mour\~ao, J. Tavares.
Complexification of gauge theories.
hep-th/9307142

\item
A.M.R. Magnon.
Self duality and CP violation in gravity.
Univ. Blaise Pascal (France) preprint (1990).

\item
J. Maluf.
Symmetry properties of Ashtekar's formulation of canonical gravity.
Universidade de Brasilia preprint, 1991.

\item
J. Maluf.
Fermi coordinates and reference frames in the ECSK theory.
SU-GP-92/1-2

\item
L.~J. Mason and J{\"o}rg Frauendiener.
 The {S}parling 3-form, {A}shtekar variables and quasi-local mass,
  1989 preprint.

\item
O. Moritsch, M. Schweda, T. Sommer, L. Tataru and H. Zerrouk.  
BRST cohomology of Yang-Mills gauge fields in the presence of gravity in 
Ashtekar variables. TUW 94-17, hep-th/9409081.

\item
N. O'Murchadha and M. Vandyck.
 Gravitational degrees of freedom in Ashtekar's formulation of
General Relativity.
 Univ. of Cork preprint - 1990

\item
P. L. Paul.
Topological Symmetries of twisted N=2 chiral supergravity in 
Ashtekar formalism.
hep-th/9504144.

\item
J. Rasmussen and M. Weis. 
Induced Topology on the hoop group. 
NBI-HE-94-46, hep-th/9410194.

\item
Carlo Rovelli and Lee Smolin.
 Loop representation for lattice gauge theory.
 1990 Pittsburgh and Syracuse preprint.

\item
C. Rovelli and L. Smolin.
Finiteness of diffeomorphism invariant operators in nonperturbative 
quantum gravity. Syracuse University preprint SU-GP-91/8-1, August 1991.

\item
M.P. Ryan, Jr.
Cosmological ``ground state'' wave functions in gravity and electromagnetism.
To appear in {\em Proceedings of VIIth Marcel Grossman Meeting on
General Relativity, 1994}
gr-qc/9312024

\item
Lee Smolin.
The Problem of Quantum Gravity: a status report (Address to the AAAS meeting,
Washington D.C., February 1991). Syracuse preprint SU-GP-91/2-1.

\item
L. Smolin.
Experimental Signatures of Quantum Gravity. To appear in
{\em The Proceedings of the 1994 Drexel Symposium on Quantum
Theory and Measurement}, World Scientific, 1997
gr-qc/9503027.

\item
L. Smolin.
Fermions and topology.
GCPG-93/9-4, gr-qc/9404010.

\item
L. Smolin and M. Varadarajan.
Degenerate solutions and the instability of the perturbative vacuum in 
nonperturbative formulations of quantum gravity.
Syracuse University preprint SU-GP-91/8-3, August 1991.

\item
T.A. Schilling.
Non-covariance of the generalized holonomies:  Examples.
CGPG-95/3-1, gr-qc/9503064.

\item
J. Schirmer.
Triad formulations of canonical gravity without a fixed reference frame.
gr-qc/9503037.

\item
C. G. Torre.
A deformation theory of self-dual Einstein spaces.
SU-GP-91/8-7, Syracuse University preprint, 1991.

\item
M. Tsuda.
General considerations of matter coupling with
selfdual connection.
gr-qc/9505019.

\item
M. Tsuda, T. Shirafuji and H.-J. K Xie.
Ashtekar Variables and Matter Coupling.
STUPP-95-138, gr-qc/9501021.

\item
R. P. Wallner.
 A new form of Einstein's equations.
 Univ of Cologne,  Germany, preprint 1990 (submitted to
Phys. Rev. Lett.)


\newpage

\section*{Recent preprints}

\item
P. Aichelburg and A. Ashtekar.
Mathematical Problems of Quantum
Gravity.
Abstracts of seminars given at the {\em Quantum
Gravity Workshop at the Erwin
Schrodinger Institute, Viena}.

\item
J. M. Aroca, H. Fort and R. Gambini
Path Integral for Lattice Staggered Fermions in
the Loop Representation. hep-lat/9607050.

\item
J.M. Aroca, H. Fort and  R. Gambini.
On the path integral loop representation of (2+1)
lattice non-Abelian theory.
hep-lat/9703007.

\item
A. Ashtekar.
Polymer Geometry at Planck Scale and Quantum Einstein Equations.
in {\em Proceedings of the 14th International Conference on General
Relativity and Gravitation}, M. Francaviglia (ed) World Scientific,
Singapore. 
hep-th/9601054.

\item
A. Ashtekar.
Geometric Issues in Quantum Gravity. To appear in {\em Geometric 
Issues in the Foundation of Science}, L. Mason et al (eds.) (Oxford
University Press). CGPG-96-61-4.

\item
A. Ashtekar.
Quantum Mechanics of Riemannian Geometry. To appear in {Proceeding 
of the Workshop on Physics and Geometry, Institut d'Estudis
Catalans, 1996}. 

\item
A. Ashtekar and A. Corichi.
Gauss Linking Number and Electro-magnetic Uncertainty Principle.
hep-th/9701136.

\item
A. Ashtekar and J. Lewandowski. Quantum Field Theory of Geometry.
to appear in {\em Proceedings,
Conference on Historical Examination and
Philosophical Reflections on Foundations
of Quantum Field Theory, Boston, MA, 1996}.
 hep-th/9603083

\item
A. Ashtekar and J. Lewandowski, D. Marolf, J. Mourao and T. Thiemann.
$SU(N)$ Quantum Yang-Mills Theory in 2-dimensions: a complete solution.
hep-th/9605128.

\item
J.C. Baez. Degenerate solutions of general relativity from topological
field theory. gr-qc/9705051.

\item
J.C. Baez and K. Krasnov.
Quantization of diffeomorphism invariant theories with Fermions.
hep-th/9703112.

\item
J.C. Baez and S. Sawin.
Functional Integration on Spaces of Connections.
q-alg/9507023.

\item
G. Barnich and V. Husain, 
Geometrical Representation of the Constraints of Euclidean General Relativity.
gr-qc/9611030.

\item
Roumen Borissov.
Graphical Evolution of Spin Netwoks States.
gr-qc/9606013.

\item
A. Corichi and J.A. Zapata.
On diffeomorphism Invariance for Lattice Theories. 
(To appear in \np{B} 1997)
gr-qc/9611034.

\item
R. De Pietri.
Spin Networks and Recoupling in Loop Quantum Gravity.
to apper in {\em Proceedings of the 2nd
Conference on Constrained Dynamics and Quantum Gravity,
Santa Margherita, Italy, 1996}.
gr-qc/9701041.

\item
C. Di Bartolo.
The Gauss constraint in the extended loop representation.
gr-qc/9607014.

\item
I. Bengtsson and A. Kleppe.
On chiral P forms.
hep-th/9609102.

\item
H. Fort, R. Gambini and J. Pullin.
Lattice Knot Theory and Quantum Gravity in the Loop
Representation. gr-qc/9608033.

\item
R. Gambini and J. Pullin.
Variational derivation of exact skein relations from 
Chern-Simons theories.
(to appear in \cmp 1997)
hep-th/9602165.

\item
H. Garcia-Compean, J. Plebanski and M. Przanowski.
Geometry associated with self-dual Yang-Mills
and the chiral model approach to selfdual gravity.
hep-th/9702046.

\item
H. Garcia-Compean, L. E. Morales and J. F. Plebanski.
A Hopf algebra structure in self-dual gravity.
CINVESTAV-FIS GRMP 10/94, hep-th/9410154.

\item
J. Griego.
On the Extended Loop Calculus.
gr-qc/9512011.

\item
Y. Hashimoto, Y. Yasui, S. Miyagi, T. Otsuka.
Applications of the Ashtekar
gravity to four dimensional hyper-K\"ahler
geometry and Yang-Mills instantons.
hep-th/9610069.

\item 
G. Immirzi.
Real and complex connections for canonical gravity.
gr-qc/9612030.

\item 
G. Immirzi.
Quantum gravity and Regge calculus.
in {\em 2nd Meeting on Constrained Dynamics and
Quantum Gravity}, Santa Margherita, Italy, 1996.
gr-qc/9701052.

\item
V. M. Khatsymovsky.
Ashtekar Constraint Surface as Projection of 
Hilbert-Palatini One.
gr-qc/9604053.

\item
K. Krasnov. On statistical mechanics of Schwarzschild black hole.
(Formerly known as: On statistical mechanics of gravitational systems).
gr-qc/9605047

\item
J. Lewandowski and J. Wisniewski.
(2+1) sectors of (3+1) gravity.
gr-qc/9609019

\item
R. Loll.
Making quantum gravity calculable.
gr-qc/9511080.

\item
R. Loll.
Further results on geometric operators in quantum
gravity.
gr-qc/9612068

\item
R. Loll.
Latticing quantum gravity.
Contributed to 2nd meeting on constrained dynamics and
quantum gravity, Santa Margherita, Italy, Sep 1996.
gr-qc/9701007

\item
R.Loll.
Quantizing canonical gravity in the real domain.
gr-qc/9701031.

\item
R. Loll.
Still on the way to quantizing gravity.
gr-qc/9701032.

\item
R.Loll
Imposing det E>0 in discrete quantum gravity.
Preprint AEI-028.
gr-qc/9703033.

\item
S. Major and L. Smolin.
Mixmaster Quantum Cosmology in Terms of Physical Dynamics.
gr-qc/9607020.

\item
F. Markopoulo and L. Smolin.
Causal evolution of spin networks.
CGPG-97/2-2.
gr-qc/9702025.

\item
D. Marolf, J. Mourao and T. Thiemann.
The status of diffeomorphism superselection in
Euclidean (2+1) gravity. (to appear in \jmp).
gr-qc/9701068.

\item
J. Pullin.
Canonical quantum gravity with new variables and loops: a report.
gr-qc/9606061.

\item
M. Reisenberger.
A Left-Handed Simplicial Action for Euclidean General Relativity.
gr-qc/9609002.

\item
M. P. Reisenberger and C. Rovelli.
`Sum Over Surfaces' Form of Loop Quantum Gravity.
gr-qc/9612035.

\item
L. Smolin.
Three dimensional strings a collective coordinates of
four-dimensional nonperturbative quantum gravity.
gr-qc/9609031. 

\item
L. Smolin.
The classical limit and the form of the Hamiltonian
constraint in nonperturbative quantum general relativity.
gr-qc/9609034.

\item
L. Smolin.
The future of spin networks.
To appear in {\em Geometric Issues in the Foundation 
of Science}, L. Mason et al (eds.) (Oxford
University Press).
qr-qc/9702030.

\item
M. Tsuda.
Consistency of matter field equations
in Ashtekar formulation.
gr-qc/9602022.

\item
M. Tsuda and T. Shirafuji.
Consistency of matter field equations in Ashtekar formulation.
gr-qc/9602022.

\item
T. Thiemann.
Quantum Spin Dynamics (QSD).
gr-qc/9606089.

\item
T. Thiemann.
Quantum Spin Dynamics II (QSD).
gr-qc/9606090.

\item
T. Thiemann.
Closed formula for the matrix elements of
the volume operator in canonical quantum gravity.
gr-qc/9606091.

\item
T. Thiemann.
A lenght operator for canonical quantum gravity.
gr-qc/9606092.

\item
T. Thiemann.
The inverse loop transform.
hep-th/9601105.

\item
T. Thiemann.
An axiomatic approach to quantum gauge field
theory.
hep-th/9511122.

\item
L.F. Urrutia.
Towards a Loop Representation of Connection
Theories Defined Over a Super-Lie Algebra.
hep-th/9609010.

\item
J.A. Zapata.
Combinatorial space from loop quantum gravity.
CGPG-97/3-8.gr-qc/9703038

\item
J.A. Zapata.
A combinatorial approach to diffeomorphism invariant
quantum gauge theories.
CGPG-97/3-7.gr-qc/9703037

\end{enumerate}

\end{document}